\documentclass[11pt,letterpaper]{article}

\usepackage[T1]{fontenc}
\usepackage[utf8]{inputenc}
\usepackage{lmodern}
\usepackage[margin=1in]{geometry}
\usepackage{graphicx}
\usepackage{amssymb,bm,mathtools,pdflscape}
\usepackage{booktabs,array,multirow,longtable}
\usepackage[dvipsnames]{xcolor}
\usepackage{microtype}
\usepackage{enumitem}
\usepackage{setspace}
\usepackage{float}
\usepackage[font=small,labelfont=bf]{caption}
\usepackage[authoryear,round,sort,longnamesfirst]{natbib}
\usepackage{hyperref}

\setlist[enumerate]{leftmargin=*,labelindent=0pt,labelsep=0.6em}
\newcounter{assumption}
\newcounter{proposition}

\newcounter{theorem}

\newcommand{\E}[1]{\mathbb{E}\{#1\}}
\newcommand{\Prb}[1]{\mathbb{P}\{#1\}}

\hypersetup{
  colorlinks=true,
  linkcolor=MidnightBlue,
  citecolor=MidnightBlue,
  urlcolor=MidnightBlue,
  pdftitle={Doubly Robust Estimation and Inference for Longitudinal Win Probabilities with Ordinal Outcomes in Observational Studies},
  pdfauthor={Shuhan Liu, Maria Kudela, Margaret Gamalo, Prosenjit Kundu},
  pdfkeywords={Causal inference, double robustness, longitudinal ordinal outcomes, targeted maximum likelihood estimation, win probability}
}

\title{Doubly Robust Estimation and Inference for Longitudinal
Win Probabilities with Ordinal Outcomes in Observational Studies}
\author{%
Shuhan Liu\textsuperscript{1}, Maria Kudela\textsuperscript{2},
Margaret Gamalo\textsuperscript{2},\\
and Prosenjit Kundu\textsuperscript{2}\thanks{Corresponding author: \href{mailto:prosenjit.kundu@pfizer.com}{\texttt{prosenjit.kundu@pfizer.com}}.}\\[0.7em]
\begin{minipage}{0.96\textwidth}
\centering\small
\textsuperscript{1}Statistics and Data Science, Northwestern University,\\
2006 Sheridan Rd, Evanston, IL 60208, US\\[0.35em]
\textsuperscript{2}Statistics, Pfizer Inc., 1 Portland St, Cambridge, MA 02169, US
\end{minipage}}
\date{}

\begin{document}
\maketitle

\begin{abstract}
Repeated ordinal outcomes capture how treatment benefits and harms evolve over time. We develop doubly robust estimation and inference for visit-specific win probabilities and their prespecified weighted summaries, with Desirability of Outcome Ranking as a motivating example. Baseline confounding is addressed through augmented inverse-probability weighting (AIPW) of treatment-specific marginal category probabilities or joint multinomial targeted maximum likelihood estimation (TMLE). We consider fold-specific and pooled targeting and establish conditions for double-robust consistency and semiparametric efficiency under participant-level cross-fitting. When only one nuisance model is correctly specified, inference based only on the efficient influence function may be invalid. For smooth parametric nuisance models, we derive covariance corrections from the complete estimating procedure, including targeting. When a correctly specified propensity model is estimated by maximum likelihood, a simplified score correction accommodates convergent flexible outcome learners without requiring their derivatives. The corrected covariance estimators retain within-participant dependence and support visit-specific, weighted-summary, and joint inference across repeated assessments. Simulations supported double robustness, demonstrated the importance of cross-fitting with overfit learners, and showed improved marginal and joint confidence coverage with the proposed corrections. TMLE, particularly with pooled targeting, often had lower root mean squared error under adequate treatment overlap, whereas AIPW was more stable under weak overlap. An electronic health record study of hospitalized adults with an acute respiratory illness illustrates how the trajectory and weighted summary reveal complementary aspects of the treatment comparison.
\end{abstract}

\noindent\textbf{Keywords:} Causal inference; double robustness;
longitudinal ordinal outcomes; targeted maximum likelihood estimation;
win probability.

\section{Introduction}
\label{sec:introduction}
Clinical decisions rarely depend on whether one event occurred by one fixed day. What matters is the patient's clinical course. In an acute illness, a patient may die, need mechanical ventilation, remain in the hospital, or recover and return home. These states have a clear clinical order, even though the difference between neighboring states is not measured in common units. Ordinal scales are therefore used in acute stroke and infectious disease research, where changes across the full range of disability or illness severity can be meaningful to patients and clinicians \citep{saver2007,marshall2020}. Timing matters too. A treatment may speed recovery early, have little effect later, or cause harms that appear only after longer exposure. A binary outcome or a single assessment at the end of follow-up can miss these differences and discard clinically useful information. This clinical problem leads directly to the statistical question addressed in this paper. How should two treatments be compared across several visits when outcomes are ordered, the spacing between categories is not defined, and treatment was not randomized? Our framework answers this question with a marginal ordinal win probability at each visit. It is the probability that an independently selected participant under one treatment has a better outcome than an independently selected participant under the other treatment, with a tie counted as one half. This comparison applies to any outcome with clinically ordered categories. Desirability of Outcome Ranking, or DOOR, is one important example in which benefit and harm are combined into a ranking of overall clinical desirability \citep{evans2015}. The visit-specific win probabilities show when treatment differences arise. Their prespecified weighted average gives a single summary across follow-up without replacing the trajectory. In observational electronic health record data, estimating these quantities also requires adjustment for baseline confounding, allowance for repeated outcomes from the same participant, and inference that captures uncertainty shared across visits.

Several strands of work provide the foundation for our approach. Finkelstein and Schoenfeld developed a pairwise procedure that combines a time-to-event outcome with a longitudinal measure. Their procedure compares survival first and uses the longitudinal outcome when the survival comparison is unresolved \citep{finkelstein1999}.
Related global-rank endpoints place each participant's overall clinical course on a common ordered scale \citep{felker2010}. Generalized pairwise comparisons compare participants according to prespecified priorities among outcomes, while the win ratio summarizes the relative number of favorable and unfavorable comparisons \citep{buyse2010,pocock2012,bebu2016}. DOOR assigns each participant to a clinically ordered benefit--risk category and summarizes the treatment comparison through the probability of a more desirable outcome \citep{evans2015}.
For a single ordered outcome, probabilistic-index models relate win probabilities to covariates \citep{thas2012}. In observational studies, causal rank methods address confounding through weighting, outcome regression, and doubly robust estimation of Mann--Whitney and more general U-statistic targets \citep{wu2014,mao2018,zhang2019,yin2024}. Augmented inverse-probability weighting (AIPW) and targeted maximum likelihood estimation (TMLE) provide the broader semiparametric foundation for combining treatment and outcome models \citep{robins1994,bang2005,vanderlaanrubin2006,vanderlaan2014}. Cross-fitting permits flexible nuisance estimation under convergence-rate conditions without requiring restrictive empirical-process assumptions \citep{kennedy2024}. In randomized trials, related methods use baseline covariates to improve precision for Mann--Whitney and other ordinal estimands, including targeted and machine-learning-based approaches \citep{vermeulen2015,diaz2016,williams2022}.

The approaches closest to our setting provide complementary building blocks. For repeated outcomes, Shu et al. compare DOOR outcomes across visits, construct simultaneous confidence bands that account for correlation over time, and propose a weighted Mann--Whitney statistic for the full follow-up period \citep{shu2025}. For observational studies, Shu et al. develop inverse-weighted, outcome-regression, and doubly robust estimators for a DOOR probability at a single assessment \citep{shu2026}. Feng et al. develop AIPW and sequential-hazard TMLE estimators for the same single-assessment setting, incorporating flexible nuisance learning, cross-fitting, and influence-function-based inference \citep{feng2026}. Together, these approaches provide the main ingredients for longitudinal causal analysis. They motivate a unified framework that combines confounding adjustment with repeated ordinal comparisons, visit-specific and overall summaries, joint inference across visits, and inference when only one working model is correctly specified.

We develop such a framework by estimating the treatment-specific marginal distribution of the ordinal outcome at each visit. These distributions provide a common basis for the visit-specific win probabilities, their prespecified weighted summary, and their covariance across visits. Our contribution has four parts. First, we define and identify a trajectory of between-participant win probabilities and its prespecified weighted summary. Second, we construct a category-level augmented inverse-probability weighted estimator, which we call category-AIPW, and a joint multinomial TMLE. Category-AIPW corrects the marginal category probabilities before comparing the two treatment distributions. The TMLE updates all category probabilities together by solving one convex optimization problem for each treatment arm, visit, and fold. This preserves nonnegative probabilities that sum to one. The construction differs algorithmically from sequential-hazard TMLE, which targets a sequence of conditional hazards through an outer iteration. Because a validated public implementation was unavailable, we present this as an algorithmic distinction rather than a claim of faster computation. Third, we establish double-robust consistency and semiparametric efficiency under participant-level cross-fitting. Fourth, we develop inference for both efficient and one-model-correct settings. When both nuisance models are consistently estimated at suitable rates, the usual efficient-influence-function covariance applies. When only one working model is correct, nuisance estimation can contribute additional first-order variation \citep{shooksa2025,wu2025}. For smooth parametric models, we derive corrected covariance formulas from the full estimation procedure, including the TMLE targeting step. The analytic formulas also explain why the usual influence-function variance can be conservative when the propensity model is correct but the outcome model is misspecified. Together, these methods provide a confounding-adjusted longitudinal treatment-comparison trajectory and an overall summary while retaining the dependence among repeated outcomes in inference.

Below, Section~\ref{sec:targets} defines and identifies the estimands. Sections~\ref{sec:estimation} and~\ref{sec:inference} develop estimation and inference, followed by simulations in Section~\ref{sec:simulations} and an electronic health record application in Section~\ref{sec:application}. Section~\ref{sec:discussion} discusses limitations and future work at last.

\section{Target estimands and identification}
\label{sec:targets}

Let \(O=\{X,A,Y(1),\ldots,Y(T)\}\sim P_0\) denote a
participant's observed record, where \(X\) contains baseline
covariates, \(A\in\{0,1\}\) denotes treatment, and
\(Y(t)\in\{1,\ldots,K\}\) is the ordinal outcome at visit \(t\),
with larger values indicating greater desirability. Participant
records are independent and identically distributed according to
\(P_0\), while repeated outcomes within a participant may be dependent. Let \(Y^a(t)\) denote
the potential outcome under treatment \(a\).

For categories \(k\) and \(\ell\), define the win score
\[
w_{k\ell}=I(k>\ell)+\tfrac12 I(k=\ell),
\]
and let \(W=(w_{k\ell})_{k,\ell=1}^K \in \mathbb{R}^{K \times K}\) denote the win-score
matrix. The visit-specific DOOR estimand compares potential
outcomes from two independently sampled participants \(i\)
and \(j\) in the population:
\begin{equation}
\psi_t
=\E{  w_{Y_i^1(t),Y_j^0(t)}  }
=\Prb{ Y_i^1(t)>Y_j^0(t) }
+\tfrac12\Prb{  Y_i^1(t)=Y_j^0(t) }.
\label{eq:target}
\end{equation}
Let \(\bm\psi=(\psi_1,\ldots,\psi_T)^\top\) denote the
dynamic DOOR trajectory. For prespecified weights
\(\omega_t\geq0\) satisfying \(\sum_{t=1}^T\omega_t=1\),
define its weighted summary as
\[
\Psi_\omega=\sum_{t=1}^T\omega_t\psi_t.
\]

The between-participant comparison in~\eqref{eq:target}
depends only on the treatment-specific marginal outcome
distributions. In contrast, a within-participant win probability
depends on the joint distribution of \(Y^1(t)\) and \(Y^0(t)\),
which is not generally identified under standard causal
assumptions (Appendix~\ref{app:proof-identification}). To identify the marginal
distributions, and hence the DOOR estimands, we impose the
following conditions.

\refstepcounter{assumption}
\paragraph{Assumption \theassumption{} (Identification)} \label{assump::identification}
For each visit \(t\), assume:
\begin{enumerate}[label=(\roman*)]
\item \emph{Consistency and no interference:}
\(Y(t)=Y^A(t)\), and one participant's treatment does not
affect another participant's outcome.
\item \emph{Conditional exchangeability:}
\(\{Y^0(t),Y^1(t)\}\perp A\mid X\).
\item \emph{Positivity:} for some \(c\in(0,1/2)\),
\(c\leq\pi(X)\leq1-c\) almost surely, where
\(\pi(x)=\Prb{  A=1\mid X=x  }\).
\end{enumerate}

Define the conditional category probabilities and their
marginal averages by
\[
q_{at}(k\mid x)=\Prb{Y(t)=k\mid A=a,X=x},
\qquad
p_{at,k}=\E{q_{at}(k\mid X)},
\]
and let \(\bm p_{at}=(p_{at,1},\ldots,p_{at,K})^\top\).

\phantomsection
\refstepcounter{proposition}
\paragraph{Proposition \theproposition{} (Identification)} \label{prop:identification}
Under Assumption~\ref{assump::identification}, \(p_{at,k}=\Prb{Y^a(t)=k}\), and
\begin{equation}
\psi_t
=\sum_{k,\ell}w_{k\ell}p_{1t,k}p_{0t,\ell}
=\bm p_{1t}^{\top}W\bm p_{0t}.
\label{eq:id}
\end{equation}
Consequently, \(\bm\psi\) and every prespecified
\(\Psi_\omega\) are identified.

\noindent
The proof is given in
Appendix~\ref{app:proof-identification}.
Equation~\eqref{eq:id} provides the basis for estimation
through the treatment-specific marginal category probabilities.
Although identification is visit-specific, joint inference
for \(\bm\psi\) in Section~\ref{sec:inference} accounts for dependence among repeated
outcomes within participants.

\section{Estimation}
\label{sec:estimation}

Building on the identification formula~\eqref{eq:id}, we
develop AIPW and TMLE estimators for the DOOR trajectory
and its weighted summaries. For TMLE, we consider both
fold-specific targeting and shared targeting across folds.
Below, we first derive the efficient influence functions that guide
estimation of the marginal category probabilities, then
construct the DOOR estimators and establish their consistency
and efficiency.

\subsection{Efficient influence functions}

Let \(\pi_a(x)=\pi(x)^a\{1-\pi(x)\}^{1-a}\).
The efficient influence function for \(p_{at,k}\) is given by
\begin{equation}
\varphi_{at,k}(O,P_0)
=
\frac{I(A=a)}{\pi_a(X)}
\{I(Y(t)=k)-q_{at}(k\mid X)\}
+q_{at}(k\mid X)-p_{at,k}.
\label{eq:eifp}
\end{equation}
Write
\(\bm{\varphi}_{at}
=(\varphi_{at,1},\ldots,\varphi_{at,K})^\top\).
By the chain rule, the efficient influence function for
\(\psi_t=\bm p_{1t}^\top W\bm p_{0t}\) is
\begin{equation}
\phi_t(O,P_0)
=
\bm{\varphi}_{1t}(O,P_0)^\top W\bm p_{0t}
+\bm p_{1t}^\top W\bm{\varphi}_{0t}(O,P_0).
\label{eq:eifpsi}
\end{equation}
The efficient influence functions for the trajectory
\(\bm\psi\) and weighted summary \(\Psi_\omega\) are therefore
\[
\bm\phi=(\phi_1,\ldots,\phi_T)^\top,
\qquad
\phi_\omega=\sum_{t=1}^T\omega_t\phi_t.
\]
Derivations are provided in Appendix~\ref{app:influence}.

\subsection{EIF-based estimators}

Two standard approaches to EIF-based estimation are augmented
inverse probability weighting (AIPW) and targeted maximum
likelihood estimation (TMLE). In this subsection, we use these approaches to construct estimators of the marginal category probabilities and combine them to obtain estimators of the DOOR trajectory and its weighted summaries.
In particular, category-AIPW adds a one-step correction to the initial marginal
estimates, whereas TMLE updates the conditional outcome
distribution to solve the corresponding empirical
influence-function equations. For TMLE, we consider both
fold-specific targeting (TMLE) and shared targeting across
folds (TMLE-P).

All methods use participant-level cross-fitting. We partition
participants into \(V\) folds \(I_1,\ldots,I_V\), independently
of their observed values, keeping each participant's repeated
outcomes in the same fold. Let \(n_v=|I_v|\) and let \(v(i)\)
denote the fold containing participant \(i\). For a generic
function \(f\), define the fold-specific empirical average
as \(P_vf=n_v^{-1}\sum_{i\in I_v}f(O_i)\).
The nuisance estimates fitted outside \(I_v\) are denoted by
\(\widehat\pi^{(-v)}\) and \(\widehat q_{at}^{(-v)}\).
Throughout, \(\widehat\pi^{(-v)}\) refers to propensity
predictions clipped to \([\varepsilon,1-\varepsilon]\),
where \(0<\varepsilon\leq c\) for \(c\) in Assumption~\ref{assump::identification}.

For each method
\(r\in\{\mathrm{AIPW},\mathrm{TMLE},\mathrm{TMLE\text{-}P}\}\),
let \(\widehat{\bm p}_{at}^{\,r,v}\) denote the vector of
marginal category estimators in fold \(v\). We pool these
vectors using fold-size weights before applying the DOOR
identification formula:
\begin{equation}
\widehat{\bm p}_{at}^{\,r}
=\sum_{v=1}^V\frac{n_v}{n}
  \widehat{\bm p}_{at}^{\,r,v},
\qquad
\widehat\psi_t^{\,r} 
=(\widehat{\bm p}_{1t}^{\,r})^\top
  W\widehat{\bm p}_{0t}^{\,r},
\qquad
\widehat\Psi_\omega^{\,r}
=\sum_{t=1}^T\omega_t\widehat\psi_t^{\,r}.
\label{eq:aipwpsi}
\end{equation}

\subsubsection{AIPW}

Applying the correction in~\eqref{eq:eifp} to the initial
estimator \(P_v\widehat q_{at}^{(-v)}(k\mid X)\) gives
\[
\widehat p_{at,k}^{\,\mathrm{AIPW},v}
=
P_v\left[
\widehat q_{at}^{(-v)}(k\mid X)
+\frac{I(A=a)}{\widehat\pi_a^{(-v)}(X)}
\{I(Y(t)=k)-\widehat q_{at}^{(-v)}(k\mid X)\}
\right],
\]
where $\widehat\pi_a^{(-v)}(x) = \{\widehat\pi^{(-v)}(x)\}^{a}
\{1-\widehat\pi^{(-v)}(x)\}^{1-a}$.
The DOOR estimators then follow from~\eqref{eq:aipwpsi}.

Here, the one-step correction is applied to the marginal
category probabilities before forming DOOR. In contrast,
a direct one-step estimator applies a correction based on
\(\phi_t\) to the initial estimator of \(\psi_t\) itself.
The two estimators differ by a product of marginal corrections
that the AIPW construction retains and the direct estimator
omits. Under the AIPW efficiency conditions, the direct
estimator has the same efficient expansion if and only if
this product is \(o_p(n^{-1/2})\) at each visit.
Appendix~\ref{app:direct-estimator} gives the exact identity and a sufficient
outcome-model rate condition for the efficiency of the direct estimator.

Although the AIPW category estimates sum to one, individual
estimates may fall outside \([0,1]\), potentially yielding
a DOOR estimate outside this range. The targeted estimators
below preserve these probability constraints.

\subsubsection{Joint multinomial TMLE}

We target the conditional category distribution through a
joint multinomial fluctuation that updates all categories
simultaneously. Given positive initial category probabilities,
define
\begin{equation}
\widehat q_{at}^{\bm\epsilon,v}(k\mid x)
=
\frac{
\widehat q_{at}^{(-v)}(k\mid x)
\exp\{\epsilon_k/\widehat\pi_a^{(-v)}(x)\}
}{
\sum_{\ell=1}^K
\widehat q_{at}^{(-v)}(\ell\mid x)
\exp\{\epsilon_\ell/\widehat\pi_a^{(-v)}(x)\}
},
\label{eq:fluct}
\end{equation}
where \(\bm\epsilon=(\epsilon_1,\ldots,\epsilon_{K-1})^\top\)
and \(\epsilon_K=0\). This multinomial logistic fluctuation
adds \(\epsilon_k/\widehat\pi_a^{(-v)}(x)\) to the initial
log odds of category \(k\) relative to category \(K\).
The normalization ensures that the updated probabilities
remain positive and sum to one.

The fluctuation parameters are estimated jointly by
minimizing the multinomial loss
\[
L_{atv}(\bm\epsilon)
=
-P_v\left[
I(A=a)\log
\widehat q_{at}^{\bm\epsilon,v}(Y(t)\mid X)
\right],
\]
which is convex in \(\bm\epsilon\).
For fold-specific targeting (TMLE), we minimize \(L_{atv}\)
separately for each \((a,t,v)\), obtaining
\(\widehat{\bm\epsilon}_{at}^{\,v}\).
For pooled targeting (TMLE-P), we minimize
\(\sum_v(n_v/n)L_{atv}\), obtaining a shared
\(\widehat{\bm\epsilon}_{at}^{\,\mathrm{TMLE\text{-}P}}\)
for each \((a,t)\).
To express both targeting schemes in common notation, define
\[
\widehat{\bm\epsilon}_{at}^{\,r,v}
=
\begin{cases}
\widehat{\bm\epsilon}_{at}^{\,v},
& r=\mathrm{TMLE},\\[3pt]
\widehat{\bm\epsilon}_{at}^{\,\mathrm{TMLE\text{-}P}},
& r=\mathrm{TMLE\text{-}P}.
\end{cases}
\]
The targeted conditional probabilities and fold-specific
marginal category estimators are
\[
\widehat q_{at}^{*,r,v}(k\mid x)
=
\widehat q_{at}^{\widehat{\bm\epsilon}_{at}^{\,r,v},v}
(k\mid x),
\qquad
\widehat p_{at,k}^{\,r,v}
=
P_v\widehat q_{at}^{*,r,v}(k\mid X).
\]
For TMLE-P, the fluctuation vector is shared across folds,
but the targeted probabilities remain fold-specific
because the initial nuisance estimates depend on \(v\).
The marginal estimators are pooled and combined
according to~\eqref{eq:aipwpsi}.

At an exact finite minimum, the multinomial score
equations imply
\[
\sum_{v=1}^V\frac{n_v}{n}P_v\left[
\frac{I(A=a)}{\widehat\pi_a^{(-v)}(X)}
\{I(Y(t)=k)-\widehat q_{at}^{*,r,v}(k\mid X)\}
\right]=0
\]
for every category \(k\). The final category equation
follows from the first \(K-1\) because the category
residuals sum to zero.
For TMLE, these equations hold separately within each
fold, while for TMLE-P, only their pooled averages need vanish.
Consequently, the pooled targeted marginal estimators
equal their augmented counterparts evaluated at the
targeted probabilities.
For both methods, the estimated category vectors remain
valid probability distributions, and the resulting
DOOR estimates lie in \([0,1]\).

This construction differs from the sequential ordinal-hazard
TMLE of \citet{feng2026}, which targets cumulative
probabilities through logistic fluctuations of conditional
hazards. Their iterative TMLE recomputes cumulative
probabilities and clever covariates after each update.
In contrast, our multinomial fluctuation directly targets
the category probabilities using fixed offsets and
covariates. All category equations are therefore solved
jointly within a single convex optimization for each arm,
visit, and targeting sample, without an outer loop that
recomputes clever covariates. Both constructions preserve
valid ordinal distributions.

\subsection{Consistency and efficiency}

Let \(\|\cdot\|_2\) denote the \(L_2(P_X)\) norm. Define
the initial nuisance errors
\[
a_{n,v}
=\|\widehat\pi^{(-v)}-\pi\|_2,
\qquad
b_{n,v}
=\max_{a,t,k}
\|\widehat q_{at}^{(-v)}(k\mid\cdot)
-q_{at}(k\mid\cdot)\|_2,
\]
and their weighted averages
$a_n=\sum_{v=1}^Va_{n,v} {n_v}/{n}$ and $
b_n=\sum_{v=1}^Vb_{n,v}{n_v}/{n}$.
Under the regularity conditions stated in
Appendix~\ref{app:est-condition}, we next establish consistency
and efficiency of the estimators defined in~\eqref{eq:aipwpsi}.

\phantomsection
\refstepcounter{theorem}
\paragraph{Theorem \thetheorem{} (Consistency and efficiency)}
\label{thm:consistency-efficiency}
Suppose Assumption~\ref{assump::identification}
and the regularity conditions (Assumption~\ref{assump:estimation} in Appendix) hold.
For each
\(r\in\{\mathrm{AIPW},\mathrm{TMLE},\mathrm{TMLE\text{-}P}\}\),
use the pooled marginal estimators in~\eqref{eq:aipwpsi}
and prespecified summary weights.
\begin{enumerate}[label=(\roman*)]
\item \textbf{Consistency.}
If
\[
\sum_{v=1}^V\frac{n_v}{n}a_{n,v}b_{n,v}=o_p(1)
\quad\text{for AIPW or TMLE},
\qquad
a_nb_n=o_p(1)
\quad\text{for TMLE-P},
\]
then, for every arm and visit,
\[
\widehat{\bm p}_{at}^{\,r}\xrightarrow{p}\bm p_{at},
\qquad
\widehat{\bm\psi}^{\,r}\xrightarrow{p}\bm\psi,
\qquad
\widehat\Psi_\omega^{\,r}\xrightarrow{p}\Psi_\omega.
\]

\item \textbf{Efficiency.}
Suppose \(a_n=o_p(1)\) and \(b_n=o_p(1)\).
Additionally, assume
\[
\sum_{v=1}^V\frac{n_v}{n}a_{n,v}b_{n,v}=o_p(n^{-1/2})
\quad\text{for AIPW or TMLE},
\quad
a_nb_n=o_p(n^{-1/2})
\quad\text{for TMLE-P}.
\]
Then
\[
\widehat{\bm p}_{at}^{\,r}-\bm p_{at}
=\frac1n\sum_{i=1}^n\bm\varphi_{at}(O_i,P_0)+o_p(n^{-1/2}).
\]
Consequently,
\[
\widehat{\bm\psi}^{\,r}-\bm\psi
=\frac1n\sum_{i=1}^n\bm\phi(O_i,P_0)
+o_p(n^{-1/2}),\quad
\widehat\Psi_\omega^{\,r}-\Psi_\omega
=\frac1n\sum_{i=1}^n\phi_\omega(O_i,P_0)
+o_p(n^{-1/2}),
\]
and the asymptotic variances equal the corresponding
semiparametric efficiency bounds. For AIPW and fold-specific
TMLE, the marginal expansion also holds within each fold,
with \(n^{-1}\sum_i\) replaced by \(P_v\).
For TMLE-P, this expansion holds only for the pooled marginal estimators.
\end{enumerate}

The proof is given in
Appendix~\ref{app:proof-theorem1}.
Under the stated conditions, the theorem establishes
double-robust consistency: consistent estimation of
either the propensity score or the conditional outcome
distribution suffices, even if the other is misspecified.
Efficiency additionally requires both nuisance components
to be consistently estimated, with the relevant product
of their errors being \(o_p(n^{-1/2})\).
Therefore, neither component must individually attain the parametric
rate, permitting flexible nuisance learners with slower
convergence rates.

For TMLE-P, the shared fluctuation couples nuisance errors
across folds, which can lead to a stronger sufficient
product-rate condition than for fold-specific TMLE.
Because its targeting equations hold only after pooling,
the theorem establishes the efficient expansion for the
pooled estimator but does not guarantee root-\(n\) convergence of the individual fold marginals.

When only one nuisance component is consistently
estimated, double-robust consistency does not by itself
justify inference based on the estimated EIF.
Nuisance estimation may contribute additional first-order
variation that such inference does not capture.
The next section develops inference procedures that
account for these contributions.

\section{Inference under nuisance-model misspecification}
\label{sec:inference}

Under the efficiency conditions in
Theorem~\ref{thm:consistency-efficiency}, inference based
on the estimated efficient influence functions is directly valid.
When only one nuisance model is correctly specified,
the estimators remain consistent under the stated
conditions, but nuisance estimation may contribute
additional first-order variation.
Subsection~\ref{subsec:complete} accounts for this variation
through the complete system of estimating equations.
For a correctly specified propensity model estimated
by maximum likelihood, we derive a simpler projection
correction in Subsection~\ref{subsec:correct}.
Finally, Subsection~\ref{subsec:inference} describes
covariance estimation and pointwise, weighted-summary,
and joint inference.

\subsection{Correction for the complete estimating procedure}
\label{subsec:complete}

We account for uncertainty from nuisance fitting and
targeting by jointly expanding their estimating equations.
For each method
\(r\in\{\mathrm{AIPW},\mathrm{TMLE},\mathrm{TMLE\text{-}P}\}\),
let \(\bm\gamma\) collect the initial nuisance parameters
across training folds. For TMLE and TMLE-P, it additionally
includes the fold-specific and shared fluctuation
parameters, respectively.
Let \(\bm s_i(\bm\gamma)\) stack participant \(i\)'s
contributions to the training-sample nuisance scores
and, for TMLE and TMLE-P, the fold-specific or pooled
targeting scores.
With the appropriate fold indicators and sample-size
weights, the fitted parameters satisfy
\[
\frac1n\sum_{i=1}^n\bm s_i(\widehat{\bm\gamma})=\bm0.
\]

Let \(\bm f_{at,i}^{\,r}(\bm\gamma)\) denote participant
\(i\)'s vector of category contributions for arm \(a\)
and visit \(t\).
Its coordinates are the augmented contributions evaluated at the initial nuisance fits
for AIPW, and the targeted category probabilities
for TMLE and TMLE-P. Thus,
\[
\widehat{\bm p}_{at}^{\,r}
=\frac1n\sum_{i=1}^n
\bm f_{at,i}^{\,r}(\widehat{\bm\gamma}).
\]
Let \(\bm\gamma^\star\) and
\(\bm p_{at}^{\,r,\star}\) denote the corresponding
probability limits.
Define the limiting score Jacobian and marginal-mean
derivative by
\[
\bm A=\lim_{n\to\infty}\frac1n\sum_{i=1}^n
P_0\!\left[\frac{\partial\bm s_i(\bm\gamma^\star)}
{\partial\bm\gamma^\top}\right],
\qquad
\bm G_{at}=\lim_{n\to\infty}\frac1n\sum_{i=1}^n
P_0\!\left[\frac{\partial\bm f_{at,i}^{\,r}(\bm\gamma^\star)}
{\partial\bm\gamma^\top}\right].
\]
Here \(P_0\) integrates over the observation with the
fold assignment held fixed.
The score and marginal contributions are detailed in
Appendix~\ref{app:inference-definitions}.

\phantomsection
\refstepcounter{proposition}
\paragraph{Proposition \theproposition{} (Corrected parametric inference)}
\label{prop:corrected-parametric}
Under Assumption~\ref{assump:inference} in
Appendix~\ref{app:inference-conditions},
\[
\widehat{\bm p}_{at}^{\,r}-\bm p_{at}^{\,r,\star}
=\frac1n\sum_{i=1}^n
\bm{\varphi}_{at,i}^{\,r,\mathrm{corr}}
+o_p(n^{-1/2}),
\]
where
\[
\bm{\varphi}_{at,i}^{\,r,\mathrm{corr}}
=\bm f_{at,i}^{\,r}(\bm\gamma^\star)
-\bm p_{at}^{\,r,\star}
-\bm G_{at}\bm A^{-1}\bm s_i(\bm\gamma^\star).
\]
Define
\(\psi_t^{\,r,\star}
=(\bm p_{1t}^{\,r,\star})^\top W\bm p_{0t}^{\,r,\star}\)
and the corrected visit-specific contribution
\[
\phi_{t,i}^{\,r,\mathrm{corr}}
=(\bm{\varphi}_{1t,i}^{\,r,\mathrm{corr}})^\top
W\bm p_{0t}^{\,r,\star}
+(\bm p_{1t}^{\,r,\star})^\top
W\bm{\varphi}_{0t,i}^{\,r,\mathrm{corr}}.
\]
Let
\(\bm\phi_i^{\,r,\mathrm{corr}}
=(\phi_{1,i}^{\,r,\mathrm{corr}},\ldots,
\phi_{T,i}^{\,r,\mathrm{corr}})^\top\),
\(\phi_{\omega,i}^{\,r,\mathrm{corr}}
=\sum_t\omega_t\phi_{t,i}^{\,r,\mathrm{corr}}\), and
\(\Psi_\omega^{\,r,\star}=\sum_t\omega_t\psi_t^{\,r,\star}\).
Then
\begin{equation}
\widehat{\bm\psi}^{\,r}-\bm\psi^{\,r,\star}
=\frac1n\sum_{i=1}^n
\bm\phi_i^{\,r,\mathrm{corr}}+o_p(n^{-1/2}),
\qquad
\widehat\Psi_\omega^{\,r}-\Psi_\omega^{\,r,\star}
=\frac1n\sum_{i=1}^n
\phi_{\omega,i}^{\,r,\mathrm{corr}}+o_p(n^{-1/2}).
\label{eq:stackedcorr}
\end{equation}
If either the limiting propensity predictions are
correct or all initial outcome predictions converge
to their true values,
\(\bm p_{at}^{\,r,\star}=\bm p_{at}\),
\(\bm\psi^{\,r,\star}=\bm\psi\), and
\(\Psi_\omega^{\,r,\star}=\Psi_\omega\).
Let
\[
\bm\Sigma_r=\lim_{n\to\infty}n^{-1}\sum_i
P_0\{\bm\phi_i^{\,r,\mathrm{corr}}
(\bm\phi_i^{\,r,\mathrm{corr}})^\top\}.
\]
Then
\[
\sqrt n(\widehat{\bm\psi}^{\,r}-\bm\psi^{\,r,\star})
\xrightarrow{d}N_T(\bm0,\bm\Sigma_r),
\qquad
\sqrt n(\widehat\Psi_\omega^{\,r}-\Psi_\omega^{\,r,\star})
\xrightarrow{d}N(0,\bm\omega^\top\bm\Sigma_r\bm\omega).
\]
The covariance estimator in~\eqref{eq:corrected-covariance},
based on the fitted corrected contributions, satisfies
\[
n\widehat V_r\xrightarrow{p}\bm\Sigma_r,
\qquad
n\bm\omega^\top\widehat V_r\bm\omega
\xrightarrow{p}\bm\omega^\top\bm\Sigma_r\bm\omega.
\]

The proof is provided in Appendix~\ref{app:proof-prop2}.
The correction accounts for additional first-order
variation from nuisance estimation that the usual
EIF-based inference may omit under misspecification.
For TMLE and TMLE-P, it also accounts for estimation
of the fluctuation parameters during targeting.
This generally requires derivatives of the complete
estimating system, rather than separate projections
onto the propensity and outcome-model scores.

\subsection{Special case of a correctly specified propensity model}
\label{subsec:correct}

When a parametric propensity model is correctly specified
and estimated by maximum likelihood, the correction
takes a simplified form that requires no derivatives
of the outcome regression and accommodates convergent flexible
outcome learners. 
Let \(\pi_{\mathrm{raw}}(x;\bm\beta)\) denote its raw prediction,
with \(\pi_{\mathrm{raw}}(x;\bm\beta_0)=\pi(x)\).
Define its likelihood score and information matrix by
\begin{equation}
\begin{aligned}
\bm s_\beta(O)
&=\left.
\frac{\partial}{\partial\bm\beta}
\left[A\log\pi_{\mathrm{raw}}(X;\bm\beta)
+(1-A)\log\{1-\pi_{\mathrm{raw}}(X;\bm\beta)\}\right]
\right|_{\bm\beta=\bm\beta_0},\\
\bm I_\beta&=P_0\{\bm s_\beta(O)\bm s_\beta(O)^\top\}.
\end{aligned}
\label{eq:propensity-score-information}
\end{equation}

For each method, let \(q_{at}^{r,\star}\) denote the
limiting initial outcome prediction for AIPW and the
limiting targeted prediction for TMLE and TMLE-P.
Define
\[
\varphi_{at,k}^{r,\star}(O)
=\frac{I(A=a)}{\pi_a(X)}
\{I(Y(t)=k)-q_{at}^{r,\star}(k\mid X)\}
+q_{at}^{r,\star}(k\mid X)-p_{at,k}.
\]
and construct \(\phi_t^{r,\star}\) and
\(\phi_\omega^{r,\star}\) using~\eqref{eq:eifpsi}
and the prespecified summary weights.
Notably, these contributions use the limiting outcome predictions
\(q_{at}^{r,\star}\), which need not equal the true
conditional probabilities \(q_{at}\).

\phantomsection
\refstepcounter{proposition}
\paragraph{Proposition \theproposition{} (Correctly specified propensity model)}
\label{prop:correct-ps}
Under Assumption~\ref{assump:correct-ps} in
Appendix~\ref{app:projection-conditions}, the propensity
model is correctly specified and estimated by maximum likelihood,
while the initial outcome predictions may converge to misspecified
limits.
For each
\(r\in\{\mathrm{AIPW},\mathrm{TMLE},\mathrm{TMLE\text{-}P}\}\),
\[
\widehat{\bm\psi}^{\,r}-\bm\psi
=\frac1n\sum_{i=1}^n
\bm\phi^{r,\mathrm{corr}}(O_i)+o_p(n^{-1/2}),
\]
where \(\bm\phi^{r,\star}
=(\phi_1^{r,\star},\ldots,\phi_T^{r,\star})^\top\) and
\[
\bm\phi^{r,\mathrm{corr}}(O_i)
=\bm\phi^{r,\star}(O_i)
-\bm C^r\bm I_\beta^{-1}\bm s_\beta(O_i),
\qquad
\bm C^r=P_0\{\bm\phi^{r,\star}(O)\bm s_\beta(O)^\top\}.
\]
For prespecified summary weights,
\[
\widehat\Psi_\omega^{\,r}-\Psi_\omega
=\frac1n\sum_{i=1}^n
\phi_\omega^{r,\mathrm{corr}}(O_i)+o_p(n^{-1/2}),
\qquad
\phi_\omega^{r,\mathrm{corr}}
=\bm\omega^\top\bm\phi^{r,\mathrm{corr}}.
\]
In particular, with \(\bm C_\omega^r=\bm\omega^\top\bm C^r\),
the corrected covariance and variance satisfy
\[
\operatorname{Var}(\bm\phi^{r,\mathrm{corr}})
=\operatorname{Var}(\bm\phi^{r,\star})
-\bm C^r\bm I_\beta^{-1}(\bm C^r)^\top
\preceq\operatorname{Var}(\bm\phi^{r,\star}),\]
\[
\operatorname{Var}(\phi_\omega^{r,\mathrm{corr}})
=\operatorname{Var}(\phi_\omega^{r,\star})
-\bm C_\omega^r\bm I_\beta^{-1}(\bm C_\omega^r)^\top
\leq\operatorname{Var}(\phi_\omega^{r,\star}),
\]
where \(\preceq\) denotes the positive semidefinite order.
Writing \(\bm\Sigma_r=\operatorname{Var}(\bm\phi^{r,\mathrm{corr}})\),
\[
\sqrt n(\widehat{\bm\psi}^{\,r}-\bm\psi)
\xrightarrow{d}N_T(\bm0,\bm\Sigma_r),
\qquad
\sqrt n(\widehat\Psi_\omega^{\,r}-\Psi_\omega)
\xrightarrow{d}N(0,\bm\omega^\top\bm\Sigma_r\bm\omega).
\]
Using the fitted score correction and consistent estimates of
\(\bm C^r\) and \(\bm I_\beta\), the covariance estimator
in~\eqref{eq:corrected-covariance} satisfies
\[
n\widehat V_r\xrightarrow{p}\bm\Sigma_r,
\qquad
n\bm\omega^\top\widehat V_r\bm\omega
\xrightarrow{p}\bm\omega^\top\bm\Sigma_r\bm\omega.
\]

The proof is provided in Appendix~\ref{app:proof-prop3}.
Several implications merit discussion.
First, the correction depends on the covariance between
the limiting augmented contribution and the propensity
likelihood score, together with the propensity information
matrix. Unlike Proposition~\ref{prop:corrected-parametric},
this result requires neither a parametric outcome model
nor derivatives or a parameter expansion for the outcome
learner. It therefore accommodates flexible learners,
provided they satisfy the stated \(L_2(P_X)\) convergence
and, for TMLE and TMLE-P, the additional targeting conditions.

Second, the variance reduction implies that uncorrected
Wald intervals can be asymptotically conservative when
the propensity model is correct but the outcome model
is misspecified. We examine this behavior in the
simulation studies.

Finally, an analogous simplification does not generally
hold when the outcome learner is consistent but the
propensity converges to an incorrect limit.
The outcome-estimation error is then multiplied by
a nonvanishing propensity error, so its first-order
contribution cannot generally be omitted, and cross-fitting alone does not eliminate this contribution.

\subsection{Covariance estimation and joint inference}
\label{subsec:inference}

Let \(\widehat{\bm\phi}_i^{\,r,\mathrm{corr}}\) denote
the fitted corrected trajectory contribution for
participant \(i\).
The covariance of \(\widehat{\bm\psi}^{\,r}\) can be estimated by
\begin{equation}
\widehat V_r
=\frac1{n(n-1)}\sum_{i=1}^n
(\widehat{\bm\phi}_i^{\,r,\mathrm{corr}}
-\bar{\bm\phi}^{\,r,\mathrm{corr}})
(\widehat{\bm\phi}_i^{\,r,\mathrm{corr}}
-\bar{\bm\phi}^{\,r,\mathrm{corr}})^\top,
\label{eq:corrected-covariance}
\end{equation}
where
\(\bar{\bm\phi}^{\,r,\mathrm{corr}}
=n^{-1}\sum_i\widehat{\bm\phi}_i^{\,r,\mathrm{corr}}\).
Under the efficiency conditions in
Theorem~\ref{thm:consistency-efficiency}, the same formula
applies with estimated efficient influence functions
in place of the corrected contributions.
Retaining all visits in each participant-level vector
accounts for within-participant dependence.

Pointwise and weighted-summary Wald intervals are
\[
\widehat\psi_t^{\,r}
\pm z_{1-\alpha/2}\sqrt{\widehat V_{r,tt}},
\qquad
\widehat\Psi_\omega^{\,r}
\pm z_{1-\alpha/2}\sqrt{\bm\omega^\top\widehat V_r\bm\omega},
\]
where \(z_{1-\alpha/2}\) is the corresponding standard
normal quantile.
These intervals have asymptotic coverage \(1-\alpha\)
when the corresponding expansions are centered at the target
estimands, the variance estimators are consistent, and the
respective limiting variances are positive.
For joint inference on the trajectory, define the
Wald confidence region
\[
\mathcal C_{1-\alpha}^{\,r}
=\left\{\bm u\in\mathbb R^T:
(\widehat{\bm\psi}^{\,r}-\bm u)^\top
\widehat V_r^{-1}
(\widehat{\bm\psi}^{\,r}-\bm u)
\leq\chi^2_{T,1-\alpha}\right\},
\]
where \(\chi^2_{T,1-\alpha}\) is the \(1-\alpha\)
quantile of the chi-square distribution with \(T\)
degrees of freedom.
When the corresponding expansion is centered at
\(\bm\psi\), asymptotic normality, covariance consistency,
and a positive definite limiting covariance give this region
asymptotic coverage \(1-\alpha\). Under Proposition~\ref{prop:corrected-parametric},
if both nuisance models are misspecified, the region instead
covers \(\bm\psi^{\,r,\star}\), which need not equal \(\bm\psi\).

\section{Simulation studies}
\label{sec:simulations}

In this section, we examined estimation and inference under correct and incorrect nuisance-model specification, assessed the role of cross-fitting for overfit outcome learners, and evaluated performance under weaker treatment overlap.

\subsection{Design}
\label{sec:simulation-design}
The primary estimand was the equally weighted average win probability across six visits. Each setting used 1,000 replicates at \(n=1{,}000\) and \(n=5{,}000\). 
Five independent baseline covariates were generated: two binary ($X_1, X_2$), one three-level categorical ($X_3$), and two normally distributed variables ($X_4, X_5$).
Treatment was generated as
\(A\mid\boldsymbol X\sim\operatorname{Bernoulli}\{\pi(\boldsymbol X)\}\),
where the logistic propensity model included all five covariates,
with all coefficients multiplied by a scale
parameter \(s\).
We used \(s=1\) as the baseline, with larger values of \(s\) moving propensity scores toward 0 or 1 and weakening treatment overlap.
Four-category potential outcomes were generated from arm-specific latent probit models, with within-participant error correlation \(0.6^{|j-k|}\) between visits \(j\) and \(k\). Monte Carlo integration over 100,000 covariate draws gave a reference average win probability of approximately 0.5858. Appendix~\ref{app:simulation} gives the detailed generating procedures and implementation details.

We compared G-computation, the doubly robust pairwise U-statistic, AIPW, and TMLE, each with and without five-fold participant-level cross-fitting, plus TMLE-P, giving nine estimators in total. Details of the G-computation and U-statistic estimators
are provided in Appendix~\ref{app:gu}. 
Outcome models were fitted separately for each treatment arm and visit. For cross-fitted AIPW and TMLE, marginal category estimates were pooled across folds before forming the DOOR contrasts, as in~\eqref{eq:aipwpsi}. Table~\ref{tab:simulation-regimes} summarizes the nuisance specifications and inference procedures for each regime, covering correct specification of both nuisance models (Regime 1), misspecification of the outcome model (Regime 2), the propensity model (Regime 3), or both models (Regime 4),
and weaker treatment overlap setting with both models correctly specified
(Regime 5). We evaluated bias, empirical standard deviation, and root mean squared error (RMSE) for the weighted summary. For AIPW and TMLE methods, we also evaluated 95\% Wald coverage for weighted summary and joint trajectory coverage using a chi-square Wald region with six degrees of freedom. Inference was not evaluated under misspecification of both models (Regime 4).

\subsection{Results}

When both models were correctly specified, all estimators had small bias and similar variability, and the confidence intervals and joint confidence regions achieved coverage close to the nominal 95\% level (Table~\ref{tab:sim-regime1}, Regime 1). Cross-fitting provided little improvement with correctly specified, low-dimensional parametric models. At \(n=1{,}000\), fold-specific TMLE had slightly greater negative bias than TMLE-P, and this difference decreased with sample size.

With the outcome model omitting \(X_4\) (Regime 2, Table~\ref{tab:sim-regimes23}), G-computation had approximately \(-24.3\%\) relative bias. AIPW, TMLE, and the U-statistic had much smaller bias with either parametric outcome learner, illustrating double robustness. Uncorrected intervals were conservative: at \(n=5{,}000\), proportional-probit AIPW and TMLE had coverage of 100.0\% and 99.2\%, versus 95.3\% and 95.5\% after correction. TMLE generally had lower RMSE than AIPW and the U-statistic. Among the cross-fitted estimators, TMLE-P had the smallest RMSE at both sample sizes.
The overfit tree illustrated the importance of cross-fitting. Without it, AIPW and TMLE retained biases of \(-0.0822\) and \(-0.0850\) at \(n=5{,}000\), with 0\% coverage even after variance correction. Cross-fitting substantially reduced bias, and corrected coverage with cross-fitting was 95.0\%, 93.9\%, and 94.8\% for AIPW-CF, TMLE-CF, and TMLE-P, respectively, with joint coverage of 94.0--94.9\%.

With a fixed incorrect propensity but a correct outcome model (Regime 3, Table~\ref{tab:sim-regimes23}), all estimators had small bias and RMSE. At \(n=5{,}000\), correction reduced weighted-summary coverage from 99.8\% to 94.8--94.9\% and joint coverage from 100\% to 93.8--94.3\%. The correction therefore improved variance calibration without changing the point estimates.

When both models omitted \(X_4\) (Table~\ref{tab:sim-regime4}), estimates remained near 0.443--0.444 rather than 0.5858. Neither cross-fitting nor larger samples removed the bias. Double robustness therefore did not protect against simultaneous misspecification of both nuisance models in this regime.

Under weaker overlap (Table~\ref{tab:sim-regime5}; Figure~\ref{fig:simulation-overlap}), fold-specific targeting was more sensitive than pooled targeting. At \(s=1.5\), TMLE-CF weighted-summary coverage was 86.6\% and 90.4\% at the two sample sizes, versus 95.0\% and 95.7\% for TMLE-P. At \(s=2\) and \(n=5{,}000\), TMLE-CF had bias \(-0.0113\), weighted-summary coverage 50.7\%, and joint coverage 55.0\%. Pooled targeting improved these to \(-0.0029\), 90.4\%, and 78.2\%, but did not restore nominal coverage. AIPW retained small bias and 95.5\% weighted-summary coverage, with joint coverage slightly below nominal. Thus, TMLE-P often offered lower RMSE under adequate overlap, whereas AIPW was more reliable under weaker overlap.

\section{Electronic health record application}
\label{sec:application}
We apply the methods to a retrospective cohort of adults, 18 years of age or older, who were hospitalized with a confirmed acute viral respiratory infection during a two-year study period. The blinded analysis compares two treatments, labeled
treated and control in the figures. The analytic cohort
and treatment-arm sample sizes are withheld in the
blinded manuscript. The estimand compares the marginal
outcome distributions under baseline initiation of the
two treatments among eligible participants.

Ordinal outcomes were assessed at days 5, 10, 15, 20,
25, and 30 using seven ordered categories based on death,
respiratory support, intensive care, hospitalization
duration, and discharge without complications.
Higher categories were more desirable, and death was
treated as absorbing. DOOR probabilities above 0.5
favor the treated group. Equal visit weights of $1/6$
were prespecified to summarize the comparison across
the six assessments.

Adjustment included 15 baseline covariates covering
demographics and clinical risk factors. Propensity
scores were estimated by logistic regression, and
proportional-probit outcome models were fitted separately
for each treatment arm and visit. Both nuisance models
included all 15 covariates as main effects, with
standardized age and categorical coding for the
remaining variables. Before adjustment, standardized
mean differences (SMDs) exceeded 0.1 for heart failure
(0.113), obesity (0.111), non-Hispanic White race and
ethnicity (0.222), and unknown race and ethnicity
(0.222), indicating mild baseline imbalance.
After inverse probability weighting using the fitted
propensity scores, all SMDs were below 0.1, suggesting
improved balance in the measured baseline covariates.
Figure~\ref{fig:application-overlap} also shows
substantial overlap between the fitted propensity-score
distributions in the two treatment groups.

Five estimators were considered: full-sample AIPW and
TMLE, cross-fitted AIPW (AIPW-CF), cross-fitted TMLE
with fold-specific targeting (TMLE-CF), and TMLE-P
with pooled targeting. 
Given the similar results across estimators and the favorable performance of TMLE-P in simulations with adequate overlap, we focus on the TMLE-P trajectory shown in Figure~\ref{fig:application-trajectory}. Complete results for all estimators are provided in Table~\ref{tab:door-real-results} in the Appendix.
Inference used the complete parametric correction in
Proposition~\ref{prop:corrected-parametric}, accounting
for nuisance fitting, targeting, and dependence across
visits. The figure displays pointwise 95\% Wald
intervals and simultaneous intervals obtained by
projecting the six-dimensional 95\% Wald confidence
region onto each visit coordinate.

At day 5, the estimated DOOR probability was 0.5950,
with pointwise 95\% CI $(0.5893,\,0.6008)$ and
simultaneous interval $(0.5846,\,0.6054)$.
At day 10, the estimate was 0.5117, with corresponding
intervals $(0.5059,\,0.5176)$ and $(0.5012,\,0.5223)$.
Both simultaneous intervals exclude 0.5, in favor of
the treated group. Estimates at days 15--30 were close
to 0.5, with both pointwise and simultaneous intervals
including 0.5. The equally weighted TMLE-P summary
was 0.5177 (95\% CI, $0.5130$--$0.5223$), with
corresponding estimates across all five methods
ranging from 0.5176 to 0.5178. Thus, the weighted
summary indicated an overall advantage for the
treated group, while the trajectory showed that
the observed differences were concentrated early
in follow-up.

\section{Discussion}
\label{sec:discussion}
Longitudinal ordinal outcomes raise complementary questions about the overall treatment comparison and its evolution over time. The prespecified weighted summary addresses the first, while the visit-specific trajectory shows whether an advantage appears early, persists, or reverses. Interpreting these together helps clarify a full treatment story than either one alone.

To estimate these complementary summaries while
adjusting for baseline confounding, we develop AIPW
and joint multinomial TMLE estimators based on
treatment-specific marginal category probabilities.
AIPW corrects the marginal category probabilities before forming the DOOR comparison, avoiding the additional convergence-rate condition required by directly correcting the DOOR plug-in estimator.
Joint multinomial TMLE offers the additional advantage
over AIPW of preserving valid category probability
distributions through a single convex targeting step
that updates all categories simultaneously.
We also address inference when one nuisance model is misspecified, where double-robust consistency alone does not ensure valid inference based on the estimated efficient influence function and nuisance estimation may contribute additional first-order variation.
The proposed complete parametric correction accounts
for this variation by jointly expanding the
nuisance-fitting and targeting equations.
When a correctly specified propensity model is estimated by maximum likelihood, the correction simplifies and accommodates convergent flexible outcome learners, without requiring their derivatives. The resulting covariance
estimators account for within-participant dependence
and support visit-specific, weighted-summary, and
joint trajectory inference.

In our simulations, TMLE, particularly TMLE-P, often achieved lower RMSE than AIPW when overlap was adequate. Under weaker overlap, AIPW was able to provide more reliable coverage, while pooled targeting reduced the bias and undercoverage observed with fold-specific TMLE.
The electronic health record application further illustrates
the complementary information provided by the trajectory
and weighted summary.
The weighted summary indicated an overall advantage for the treated group, while the trajectory showed that the observed differences were concentrated at days 5 and 10. 
However, these findings do not remove the limitations of observational electronic health record evidence. Treatment-version consistency, outcome ascertainment, center-level dependence, residual confounding, and the clinical justification for the outcome ranking all warrant transparent sensitivity analyses.

Several directions remain open for future research.
The current framework compares treatment-specific
marginal outcome distributions at repeated visits
under a baseline treatment. It does not rank entire
outcome trajectories or address time-varying treatment
strategies or mediation through postbaseline covariates.
Future work could accommodate intermittent missingness
and informative censoring, extend estimation and
inference to time-varying treatments and confounding,
and incorporate data-adaptive visit weights, with
inference accounting for the uncertainty introduced
by weight selection.
A direct comparison of joint multinomial targeting
with sequential-hazard targeting also remains open.
Such a benchmark study should use validated implementations
in a common software environment and evaluate bias,
root mean squared error, coverage, convergence failures,
and running time across a range of data-generating
settings. Although joint multinomial targeting uses
a single convex optimization, this construction alone
does not establish empirical computational superiority.
A systematic comparison is needed to determine whether
it offers practical advantages in numerical stability
or computational cost.

\section*{Acknowledgments}
\noindent
\textit{Funding.} This research received no external grant funding. SL worked on this research during a paid internship at Pfizer Inc.

\noindent
\textit{Conflicts of interest.} Maria Kudela, Margaret Gamalo, and Prosenjit Kundu are employees and shareholders of Pfizer Inc. The authors declare no further conflicts of interest. 

\noindent
\textit{AI assistance.} OpenAI ChatGPT and Codex were used for language editing, manuscript organization, coding assistance, and assistance with reviewing mathematical derivations. The authors reviewed the AI-assisted material and retain responsibility for the methods, analyses, and conclusions.

\noindent
\section*{Data and Code Availability}
The data are licensed and cannot be redistributed by the authors. R code implementing the proposed methods is available on GitHub at \url{https://github.com/DynamicDOOR/DynamicDOOR}.

\section*{Appendices and code}
Appendices A--E are included below and provide proofs, implementation details, and complete application results. R code is available from the repository listed in the Data and Code Availability statement.
\clearpage
\begingroup
\singlespacing
\section*{Tables}

\begin{table}[!ht]
\centering
\begingroup
\singlespacing
\small
\setlength{\tabcolsep}{3pt}
\renewcommand{\arraystretch}{1.08}
\caption{Simulation regimes.}
\label{tab:simulation-regimes}
\begin{tabular}{@{}
>{\raggedright\arraybackslash}p{0.08\linewidth}
>{\raggedright\arraybackslash}p{0.20\linewidth}
>{\raggedright\arraybackslash}p{0.23\linewidth}
>{\raggedright\arraybackslash}p{0.25\linewidth}
>{\raggedright\arraybackslash}p{0.18\linewidth}@{}}
\toprule
Regime & Propensity model & Outcome model & Inference & Purpose\\
\midrule
1 & Correct logistic & Correct probit & Uncorrected EIF & Reference\\
\addlinespace[4pt]
2 & Correct logistic & Probit, multinomial, tree; omit \(X_4\)
  & Uncorrected and propensity-score-corrected
    (Proposition~\ref{prop:correct-ps})
  & Outcome misspecification\\
\addlinespace[4pt]
3 & Fixed probability 0.2 & Correct probit
  & Uncorrected and parametrically corrected
    (Proposition~\ref{prop:corrected-parametric})
  & Propensity misspecification\\
\addlinespace[4pt]
4 & Logistic; omit \(X_4\) & Probit, multinomial, tree; omit \(X_4\)
  & Point estimation only & Both models misspecified\\
\addlinespace[4pt]
5 & Correct logistic & Correct probit & Uncorrected EIF
  & Weaker overlap: \(s\in\{1.5,2\}\)\\
\bottomrule
\end{tabular}
\par\smallskip
\begin{minipage}{\linewidth}
\footnotesize
Probit denotes proportional-probit regression; multinomial denotes
multinomial logistic regression; tree denotes a deliberately overfit
classification tree. Correct models include all five covariates.
\end{minipage}
\endgroup
\end{table}
\clearpage

\fontsize{9}{10.5}\selectfont
\setlength{\tabcolsep}{3pt}
\renewcommand{\arraystretch}{1.05}
\setlength{\LTcapwidth}{\linewidth}
\setlength{\LTpost}{6pt}

\begin{longtable}{@{}lrrrrr@{\hspace{14pt}}rrrrr@{}}
\caption{Regime 1: both nuisance models correctly specified.}
\label{tab:sim-regime1}\\
\toprule
Estimator
& \multicolumn{5}{c}{$n=1{,}000$}
& \multicolumn{5}{c}{$n=5{,}000$}\\
\cmidrule(lr){2-6}\cmidrule(lr){7-11}
& Bias & SD & RMSE & CP & JCP
& Bias & SD & RMSE & CP & JCP\\
\midrule
\endfirsthead

\multicolumn{11}{l}{\tablename\ \thetable{} (continued)}\\
\toprule
Estimator
& \multicolumn{5}{c}{$n=1{,}000$}
& \multicolumn{5}{c}{$n=5{,}000$}\\
\cmidrule(lr){2-6}\cmidrule(lr){7-11}
& Bias & SD & RMSE & CP & JCP
& Bias & SD & RMSE & CP & JCP\\
\midrule
\endhead

\midrule
\multicolumn{11}{r}{\footnotesize Continued on next page}\\
\endfoot

\bottomrule
\endlastfoot

\multicolumn{11}{l}{\textit{Regime 1: both nuisance models correctly specified}}\\*
G-comp    & $-0.0011$ & 0.0100 & 0.0101 & -- & -- & $-0.0006$ & 0.0045 & 0.0045 & -- & --\\*
G-comp-CF & $-0.0011$ & 0.0100 & 0.0101 & -- & -- & $-0.0006$ & 0.0045 & 0.0045 & -- & --\\*
U-stat    & $-0.0009$ & 0.0102 & 0.0103 & -- & -- & $-0.0006$ & 0.0045 & 0.0046 & -- & --\\*
U-stat-CF & $-0.0009$ & 0.0104 & 0.0104 & -- & -- & $-0.0006$ & 0.0045 & 0.0046 & -- & --\\*
AIPW      & $-0.0009$ & 0.0102 & 0.0103 & 95.3 & 95.1 & $-0.0006$ & 0.0045 & 0.0046 & 96.0 & 94.3\\*
AIPW-CF   & $-0.0009$ & 0.0103 & 0.0104 & 95.5 & 96.4 & $-0.0006$ & 0.0045 & 0.0046 & 96.1 & 94.8\\*
TMLE      & $-0.0009$ & 0.0102 & 0.0102 & 95.2 & 94.9 & $-0.0006$ & 0.0045 & 0.0046 & 96.0 & 94.1\\*
TMLE-CF   & $-0.0018$ & 0.0104 & 0.0106 & 94.4 & 95.0 & $-0.0007$ & 0.0045 & 0.0046 & 96.8 & 94.6\\*
TMLE-P    & $-0.0009$ & 0.0103 & 0.0103 & 95.4 & 96.4 & $-0.0006$ & 0.0045 & 0.0046 & 96.2 & 94.6\\

\end{longtable}

{\footnotesize\noindent
Bias, empirical standard deviation (SD), and root mean squared error
(RMSE) refer to the equally weighted average win probability.
CP and JCP denote coverage (\%) of its nominal 95\% Wald interval and
the nominal 95\% joint chi-square Wald region for all six visits, respectively.
CF denotes five-fold cross-fitting.
\par}

\clearpage
\begin{longtable}{@{}lrrrrr@{\hspace{14pt}}rrrrr@{}}
\caption{Regimes 2 and 3: one nuisance model correctly specified.}
\label{tab:sim-regimes23}\\
\toprule
Estimator
& \multicolumn{5}{c}{$n=1{,}000$}
& \multicolumn{5}{c}{$n=5{,}000$}\\
\cmidrule(lr){2-6}\cmidrule(lr){7-11}
& Bias & SD & RMSE & CP & JCP
& Bias & SD & RMSE & CP & JCP\\
\midrule
\endfirsthead

\multicolumn{11}{l}{\tablename\ \thetable{} (continued)}\\
\toprule
Estimator
& \multicolumn{5}{c}{$n=1{,}000$}
& \multicolumn{5}{c}{$n=5{,}000$}\\
\cmidrule(lr){2-6}\cmidrule(lr){7-11}
& Bias & SD & RMSE & CP & JCP
& Bias & SD & RMSE & CP & JCP\\
\midrule
\endhead

\midrule
\multicolumn{11}{r}{\footnotesize Continued on next page}\\
\endfoot

\bottomrule
\endlastfoot

\multicolumn{11}{l}{\textit{Regime 2: correct propensity; proportional-probit outcome regression omitting \(X_4\)}}\\*
G-comp    & $-0.1424$ & 0.0148 & 0.1432 & -- & -- & $-0.1422$ & 0.0065 & 0.1424 & -- & --\\*
G-comp-CF & $-0.1425$ & 0.0148 & 0.1432 & -- & -- & $-0.1422$ & 0.0065 & 0.1424 & -- & --\\*
U-stat    & $-0.0017$ & 0.0127 & 0.0128 & -- & -- & $-0.0008$ & 0.0058 & 0.0058 & -- & --\\*
U-stat-CF & $0.0093$ & 0.0146 & 0.0173 & -- & -- & $0.0013$ & 0.0059 & 0.0061 & -- & --\\*
AIPW      & $-0.0017$ & 0.0121 & 0.0122 & 99.9/94.5 & 97.1/95.0 & $-0.0008$ & 0.0055 & 0.0055 & 100.0/95.3 & 96.2/94.2\\*
AIPW-CF   & $0.0069$ & 0.0130 & 0.0147 & 99.8/94.5 & 97.5/95.8 & $0.0009$ & 0.0056 & 0.0056 & 100.0/95.7 & 96.2/94.4\\*
TMLE      & $-0.0015$ & 0.0104 & 0.0105 & 98.7/94.8 & 96.4/94.8 & $-0.0007$ & 0.0046 & 0.0047 & 99.2/95.5 & 96.0/94.4\\*
TMLE-CF   & $-0.0050$ & 0.0110 & 0.0120 & 97.4/93.5 & 95.7/94.2 & $-0.0014$ & 0.0046 & 0.0048 & 99.0/95.3 & 95.5/94.4\\*
TMLE-P    & $-0.0004$ & 0.0105 & 0.0105 & 98.9/95.4 & 97.6/97.1 & $-0.0005$ & 0.0047 & 0.0047 & 99.3/95.5 & 96.3/94.8\\

\addlinespace[5pt]
\multicolumn{11}{l}{\textit{Regime 2: correct propensity; multinomial logistic outcome regression omitting \(X_4\)}}\\*
G-comp    & $-0.1425$ & 0.0149 & 0.1432 & -- & -- & $-0.1422$ & 0.0065 & 0.1424 & -- & --\\*
G-comp-CF & $-0.1425$ & 0.0149 & 0.1433 & -- & -- & $-0.1422$ & 0.0065 & 0.1424 & -- & --\\*
U-stat    & $-0.0020$ & 0.0127 & 0.0129 & -- & -- & $-0.0008$ & 0.0058 & 0.0058 & -- & --\\*
U-stat-CF & $0.0093$ & 0.0146 & 0.0173 & -- & -- & $0.0013$ & 0.0059 & 0.0061 & -- & --\\*
AIPW      & $-0.0019$ & 0.0121 & 0.0122 & 99.9/94.5 & 97.2/95.0 & $-0.0008$ & 0.0055 & 0.0056 & 100.0/95.3 & 96.2/94.2\\*
AIPW-CF   & $0.0069$ & 0.0130 & 0.0147 & 99.9/94.3 & 97.3/95.5 & $0.0009$ & 0.0056 & 0.0056 & 100.0/95.9 & 96.3/94.2\\*
TMLE      & $-0.0018$ & 0.0104 & 0.0106 & 98.8/95.1 & 96.5/94.8 & $-0.0008$ & 0.0046 & 0.0047 & 99.1/95.4 & 95.9/94.3\\*
TMLE-CF   & $-0.0050$ & 0.0110 & 0.0121 & 97.2/93.6 & 95.7/94.5 & $-0.0014$ & 0.0046 & 0.0048 & 99.0/95.2 & 95.7/94.6\\*
TMLE-P    & $-0.0004$ & 0.0106 & 0.0106 & 99.0/95.2 & 97.5/97.2 & $-0.0005$ & 0.0047 & 0.0047 & 99.3/95.4 & 96.2/94.7\\

\addlinespace[5pt]
\multicolumn{11}{l}{\textit{Regime 2: correct propensity; classification tree omitting \(X_4\)}}\\*
G-comp    & $-0.1428$ & 0.0154 & 0.1436 & -- & -- & $-0.1422$ & 0.0069 & 0.1423 & -- & --\\*
G-comp-CF & $-0.1419$ & 0.0153 & 0.1427 & -- & -- & $-0.1416$ & 0.0070 & 0.1418 & -- & --\\*
U-stat    & $-0.0850$ & 0.0138 & 0.0861 & -- & -- & $-0.0834$ & 0.0060 & 0.0836 & -- & --\\*
U-stat-CF & $0.0090$ & 0.0176 & 0.0198 & -- & -- & $0.0012$ & 0.0069 & 0.0070 & -- & --\\*
AIPW      & $-0.0838$ & 0.0137 & 0.0849 & 0.0/0.0 & 0.0/0.0 & $-0.0822$ & 0.0060 & 0.0824 & 0.0/0.0 & 0.0/0.0\\*
AIPW-CF   & $0.0066$ & 0.0160 & 0.0173 & 99.6/94.1 & 96.2/95.4 & $0.0008$ & 0.0066 & 0.0066 & 99.8/95.0 & 95.7/94.4\\*
TMLE      & $-0.0868$ & 0.0135 & 0.0879 & 0.0/0.0 & 0.0/0.0 & $-0.0850$ & 0.0060 & 0.0852 & 0.0/0.0 & 0.0/0.0\\*
TMLE-CF   & $-0.0067$ & 0.0133 & 0.0149 & 97.4/92.2 & 94.2/93.3 & $-0.0015$ & 0.0058 & 0.0060 & 97.9/93.9 & 96.1/94.9\\*
TMLE-P    & $0.0002$ & 0.0132 & 0.0132 & 99.0/95.7 & 95.8/95.4 & $-0.0004$ & 0.0058 & 0.0059 & 98.7/94.8 & 95.6/94.0\\

\addlinespace[5pt]
\multicolumn{11}{l}{\textit{Regime 3: correct proportional-probit outcome model; fixed working propensity 0.2}}\\*
G-comp    & $-0.0011$ & 0.0100 & 0.0101 & -- & -- & $-0.0006$ & 0.0045 & 0.0045 & -- & --\\*
G-comp-CF & $-0.0011$ & 0.0100 & 0.0101 & -- & -- & $-0.0006$ & 0.0045 & 0.0045 & -- & --\\*
U-stat    & $-0.0011$ & 0.0101 & 0.0102 & -- & -- & $-0.0006$ & 0.0045 & 0.0046 & -- & --\\*
U-stat-CF & $-0.0011$ & 0.0102 & 0.0103 & -- & -- & $-0.0006$ & 0.0045 & 0.0046 & -- & --\\*
AIPW      & $-0.0011$ & 0.0101 & 0.0102 & 99.9/94.6 & 100.0/95.1 & $-0.0006$ & 0.0045 & 0.0046 & 99.8/94.8 & 100.0/94.1\\*
AIPW-CF   & $-0.0011$ & 0.0102 & 0.0102 & 99.9/95.2 & 100.0/96.3 & $-0.0006$ & 0.0046 & 0.0046 & 99.8/94.9 & 100.0/94.3\\*
TMLE      & $-0.0011$ & 0.0101 & 0.0101 & 99.8/94.7 & 100.0/95.3 & $-0.0006$ & 0.0045 & 0.0045 & 99.8/94.8 & 100.0/93.8\\*
TMLE-CF   & $-0.0016$ & 0.0101 & 0.0102 & 99.8/94.5 & 100.0/95.3 & $-0.0007$ & 0.0045 & 0.0045 & 99.8/94.9 & 100.0/94.1\\*
TMLE-P    & $-0.0011$ & 0.0101 & 0.0101 & 99.9/94.7 & 100.0/95.9 & $-0.0006$ & 0.0045 & 0.0045 & 99.8/94.8 & 100.0/94.0\\
\end{longtable}

{\footnotesize\noindent
Bias, empirical standard deviation (SD), and root mean squared error
(RMSE) refer to the equally weighted average win probability.
CP and JCP denote coverage (\%) of its nominal 95\% Wald interval and
the nominal 95\% joint chi-square Wald region for all six visits, respectively.
Coverage entries are uncorrected/corrected.
CF denotes five-fold cross-fitting.
\par}

\clearpage
\begin{longtable}{@{}lrrr@{\hspace{20pt}}rrr@{}}
\caption{Regime 4: both nuisance models misspecified.}
\label{tab:sim-regime4}\\
\toprule
Estimator
& \multicolumn{3}{c}{$n=1{,}000$}
& \multicolumn{3}{c}{$n=5{,}000$}\\
\cmidrule(lr){2-4}\cmidrule(lr){5-7}
& Bias & SD & RMSE & Bias & SD & RMSE\\
\midrule
\endfirsthead

\multicolumn{7}{l}{\tablename\ \thetable{} (continued)}\\
\toprule
Estimator
& \multicolumn{3}{c}{$n=1{,}000$}
& \multicolumn{3}{c}{$n=5{,}000$}\\
\cmidrule(lr){2-4}\cmidrule(lr){5-7}
& Bias & SD & RMSE & Bias & SD & RMSE\\
\midrule
\endhead

\midrule
\multicolumn{7}{r}{\footnotesize Continued on next page}\\
\endfoot

\bottomrule
\endlastfoot

\multicolumn{7}{l}{\textit{Proportional-probit regression}}\\*
G-comp    & $-0.1424$ & 0.0148 & 0.1432 & $-0.1422$ & 0.0065 & 0.1424\\*
G-comp-CF & $-0.1425$ & 0.0148 & 0.1432 & $-0.1422$ & 0.0065 & 0.1424\\*
U-stat    & $-0.1425$ & 0.0150 & 0.1433 & $-0.1423$ & 0.0066 & 0.1425\\*
U-stat-CF & $-0.1425$ & 0.0151 & 0.1433 & $-0.1423$ & 0.0066 & 0.1425\\*
AIPW      & $-0.1425$ & 0.0150 & 0.1433 & $-0.1423$ & 0.0066 & 0.1425\\*
AIPW-CF   & $-0.1425$ & 0.0151 & 0.1433 & $-0.1423$ & 0.0066 & 0.1425\\*
TMLE      & $-0.1425$ & 0.0150 & 0.1433 & $-0.1423$ & 0.0066 & 0.1424\\*
TMLE-CF   & $-0.1424$ & 0.0150 & 0.1432 & $-0.1423$ & 0.0066 & 0.1424\\*
TMLE-P    & $-0.1425$ & 0.0150 & 0.1433 & $-0.1423$ & 0.0066 & 0.1425\\

\addlinespace[5pt]
\multicolumn{7}{l}{\textit{Multinomial logistic regression}}\\*
G-comp    & $-0.1425$ & 0.0149 & 0.1432 & $-0.1422$ & 0.0065 & 0.1424\\*
G-comp-CF & $-0.1425$ & 0.0149 & 0.1433 & $-0.1422$ & 0.0065 & 0.1424\\*
U-stat    & $-0.1425$ & 0.0150 & 0.1433 & $-0.1423$ & 0.0066 & 0.1424\\*
U-stat-CF & $-0.1425$ & 0.0151 & 0.1433 & $-0.1423$ & 0.0066 & 0.1425\\*
AIPW      & $-0.1425$ & 0.0150 & 0.1433 & $-0.1423$ & 0.0066 & 0.1424\\*
AIPW-CF   & $-0.1425$ & 0.0151 & 0.1433 & $-0.1423$ & 0.0066 & 0.1425\\*
TMLE      & $-0.1425$ & 0.0150 & 0.1433 & $-0.1423$ & 0.0066 & 0.1424\\*
TMLE-CF   & $-0.1424$ & 0.0150 & 0.1432 & $-0.1423$ & 0.0066 & 0.1424\\*
TMLE-P    & $-0.1425$ & 0.0151 & 0.1433 & $-0.1423$ & 0.0066 & 0.1425\\

\addlinespace[5pt]
\multicolumn{7}{l}{\textit{Classification tree}}\\*
G-comp    & $-0.1428$ & 0.0154 & 0.1436 & $-0.1422$ & 0.0069 & 0.1423\\*
G-comp-CF & $-0.1419$ & 0.0153 & 0.1427 & $-0.1416$ & 0.0070 & 0.1418\\*
U-stat    & $-0.1425$ & 0.0155 & 0.1434 & $-0.1420$ & 0.0069 & 0.1422\\*
U-stat-CF & $-0.1426$ & 0.0162 & 0.1435 & $-0.1423$ & 0.0071 & 0.1425\\*
AIPW      & $-0.1425$ & 0.0155 & 0.1434 & $-0.1420$ & 0.0069 & 0.1422\\*
AIPW-CF   & $-0.1426$ & 0.0162 & 0.1435 & $-0.1423$ & 0.0071 & 0.1425\\*
TMLE      & $-0.1425$ & 0.0155 & 0.1434 & $-0.1420$ & 0.0069 & 0.1422\\*
TMLE-CF   & $-0.1423$ & 0.0162 & 0.1432 & $-0.1422$ & 0.0071 & 0.1424\\*
TMLE-P    & $-0.1425$ & 0.0161 & 0.1435 & $-0.1423$ & 0.0071 & 0.1425\\
\end{longtable}

{\footnotesize\noindent
Bias, empirical standard deviation (SD), and root mean squared error
(RMSE) refer to the equally weighted average win probability.
CF denotes five-fold cross-fitting.
\par}

\clearpage
\begin{longtable}{@{}lrrrrr@{\hspace{14pt}}rrrrr@{}}
\caption{Regime 5: weaker treatment overlap.}
\label{tab:sim-regime5}\\
\toprule
Estimator
& \multicolumn{5}{c}{$n=1{,}000$}
& \multicolumn{5}{c}{$n=5{,}000$}\\
\cmidrule(lr){2-6}\cmidrule(lr){7-11}
& Bias & SD & RMSE & CP & JCP
& Bias & SD & RMSE & CP & JCP\\
\midrule
\endfirsthead

\multicolumn{11}{l}{\tablename\ \thetable{} (continued)}\\
\toprule
Estimator
& \multicolumn{5}{c}{$n=1{,}000$}
& \multicolumn{5}{c}{$n=5{,}000$}\\
\cmidrule(lr){2-6}\cmidrule(lr){7-11}
& Bias & SD & RMSE & CP & JCP
& Bias & SD & RMSE & CP & JCP\\
\midrule
\endhead

\midrule
\multicolumn{11}{r}{\footnotesize Continued on next page}\\
\endfoot

\bottomrule
\endlastfoot

\multicolumn{11}{l}{\textit{Propensity scale 1.5}}\\*
G-comp    & $-0.0013$ & 0.0104 & 0.0105 & -- & -- & $-0.0006$ & 0.0046 & 0.0046 & -- & --\\*
G-comp-CF & $-0.0013$ & 0.0104 & 0.0105 & -- & -- & $-0.0006$ & 0.0046 & 0.0046 & -- & --\\*
U-stat    & $-0.0010$ & 0.0115 & 0.0116 & -- & -- & $-0.0007$ & 0.0049 & 0.0050 & -- & --\\*
U-stat-CF & $-0.0008$ & 0.0122 & 0.0122 & -- & -- & $-0.0007$ & 0.0049 & 0.0050 & -- & --\\*
AIPW      & $-0.0009$ & 0.0115 & 0.0116 & 94.6 & 94.5 & $-0.0007$ & 0.0049 & 0.0050 & 95.9 & 93.7\\*
AIPW-CF   & $-0.0009$ & 0.0119 & 0.0120 & 94.9 & 95.2 & $-0.0007$ & 0.0049 & 0.0050 & 96.0 & 93.9\\*
TMLE      & $-0.0019$ & 0.0115 & 0.0117 & 94.1 & 92.0 & $-0.0008$ & 0.0049 & 0.0050 & 95.4 & 92.4\\*
TMLE-CF   & $-0.0097$ & 0.0120 & 0.0155 & 86.6 & 86.2 & $-0.0033$ & 0.0051 & 0.0061 & 90.4 & 90.4\\*
TMLE-P    & $-0.0019$ & 0.0116 & 0.0118 & 95.0 & 92.9 & $-0.0008$ & 0.0049 & 0.0050 & 95.7 & 92.9\\

\addlinespace[5pt]
\multicolumn{11}{l}{\textit{Propensity scale 2}}\\*
G-comp    & $-0.0016$ & 0.0108 & 0.0110 & -- & -- & $-0.0006$ & 0.0047 & 0.0047 & -- & --\\*
G-comp-CF & $-0.0017$ & 0.0108 & 0.0110 & -- & -- & $-0.0006$ & 0.0047 & 0.0047 & -- & --\\*
U-stat    & $-0.0013$ & 0.0141 & 0.0141 & -- & -- & $-0.0006$ & 0.0059 & 0.0059 & -- & --\\*
U-stat-CF & $-0.0012$ & 0.0165 & 0.0165 & -- & -- & $-0.0005$ & 0.0061 & 0.0061 & -- & --\\*
AIPW      & $-0.0013$ & 0.0141 & 0.0141 & 94.4 & 92.1 & $-0.0006$ & 0.0059 & 0.0059 & 95.5 & 92.9\\*
AIPW-CF   & $-0.0013$ & 0.0157 & 0.0158 & 94.7 & 93.1 & $-0.0005$ & 0.0060 & 0.0060 & 95.5 & 93.3\\*
TMLE      & $-0.0081$ & 0.0150 & 0.0170 & 85.0 & 71.7 & $-0.0029$ & 0.0064 & 0.0070 & 89.7 & 76.9\\*
TMLE-CF$^\dagger$ & $-0.0256$ & 0.0151 & 0.0297 & 47.9 & 52.6 & $-0.0113$ & 0.0065 & 0.0130 & 50.7 & 55.0\\*
TMLE-P    & $-0.0080$ & 0.0151 & 0.0171 & 87.0 & 76.7 & $-0.0029$ & 0.0064 & 0.0070 & 90.4 & 78.2\\
\end{longtable}

{\footnotesize\noindent
Bias, empirical standard deviation (SD), and root mean squared error
(RMSE) refer to the equally weighted average win probability.
CP and JCP denote coverage (\%) of its nominal 95\% Wald interval and
the nominal 95\% joint chi-square Wald region for all six visits, respectively.
CF denotes five-fold cross-fitting.
$^\dagger$999 valid replicates at \(s=2\), \(n=1{,}000\).
\par}

\endgroup
\clearpage
\begingroup
\singlespacing
\begin{landscape}

\section*{Figures}
\begin{figure}[H]
\centering
\includegraphics[width=\linewidth]{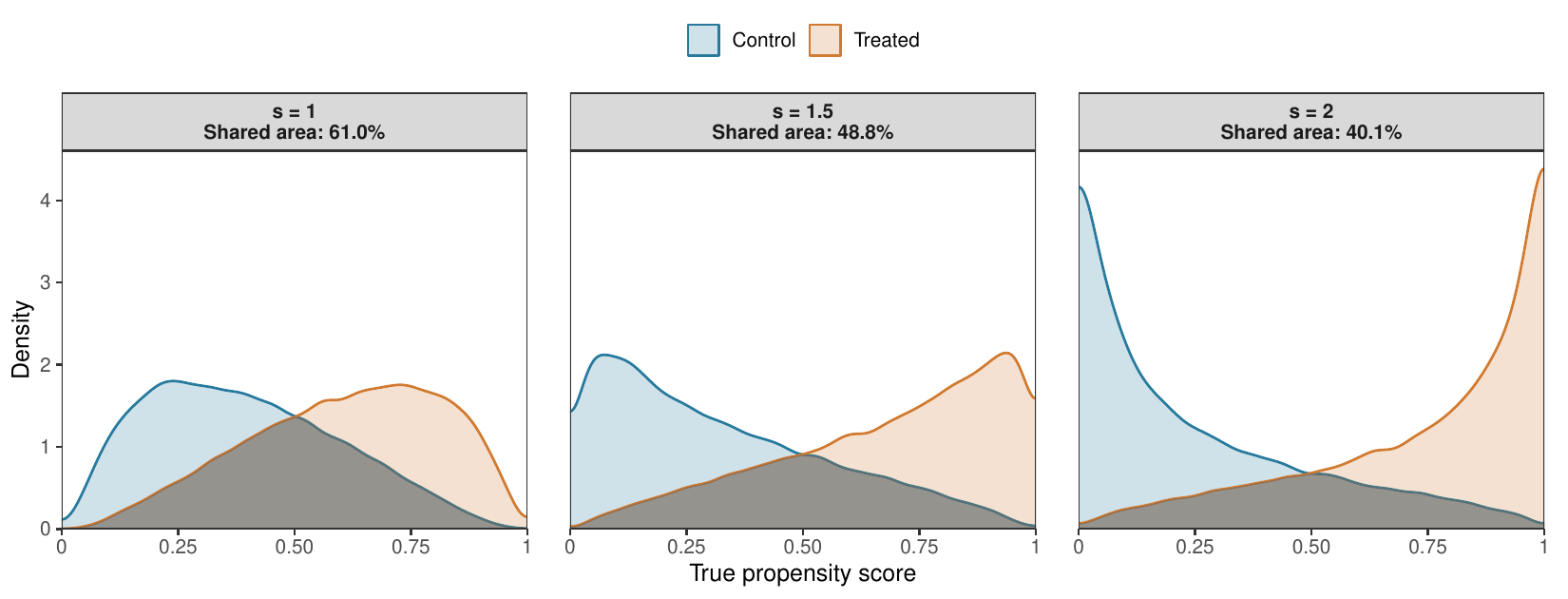}
\caption{Treatment overlap for \(s=1,1.5,2\).}
\label{fig:simulation-overlap}
\par\smallskip
{\small\raggedright
\textit{Alt text:} Three density panels show decreasing
treatment overlap as \(s\) increases from 1 to 1.5 to 2.
Control scores concentrate toward zero and treated scores
toward one. The shared areas decrease from 61.0\% to
48.8\% and 40.1\%, respectively.\par}
\end{figure}

\end{landscape}

\begin{figure}[!ht]
    \centering
    \includegraphics[width=\linewidth]{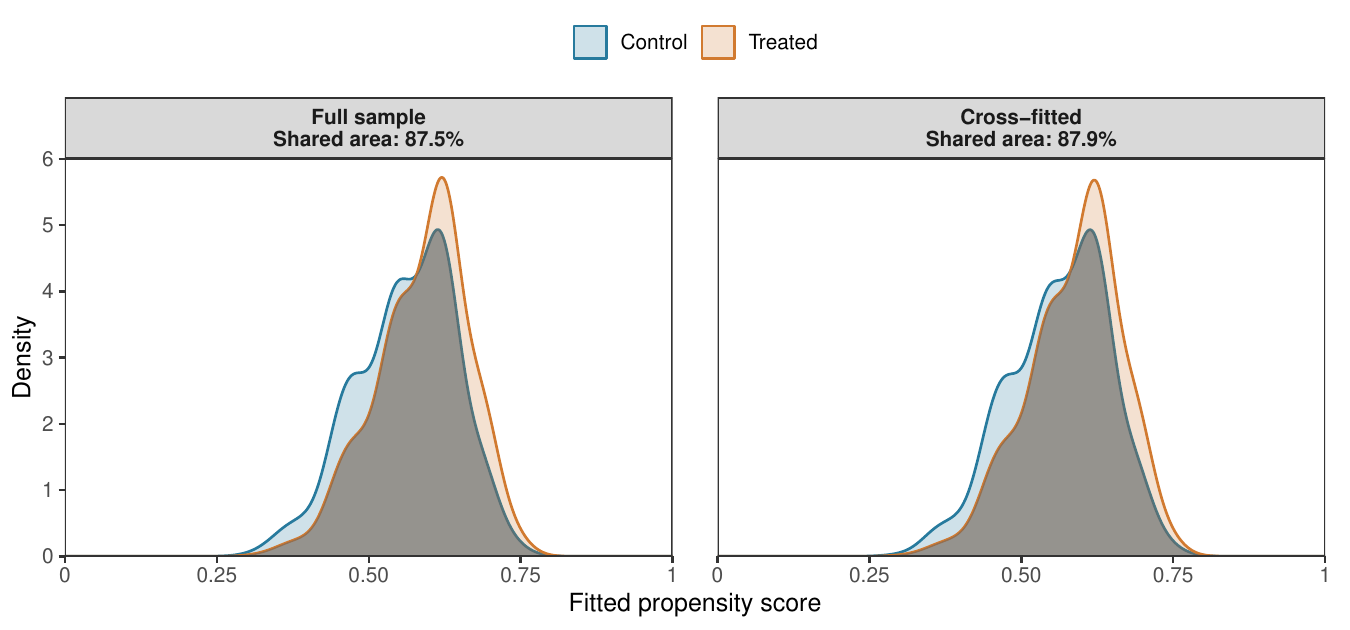}
    \caption{Fitted propensity-score distributions by treatment group in the data. The left panel uses full-sample logistic regression; the right panel uses out-of-fold predictions from five-fold cross-fitting.}
    \label{fig:application-overlap}
\par\smallskip
{\small\raggedright\textit{Alt text:} Two density panels show substantial overlap between the control and treated groups. The shared area is 87.5\% for full-sample fits and 87.9\% for cross-fitted predictions.\par}
\end{figure}
\clearpage

\begin{figure}[!ht]
    \centering
    \includegraphics[width=0.85\linewidth]{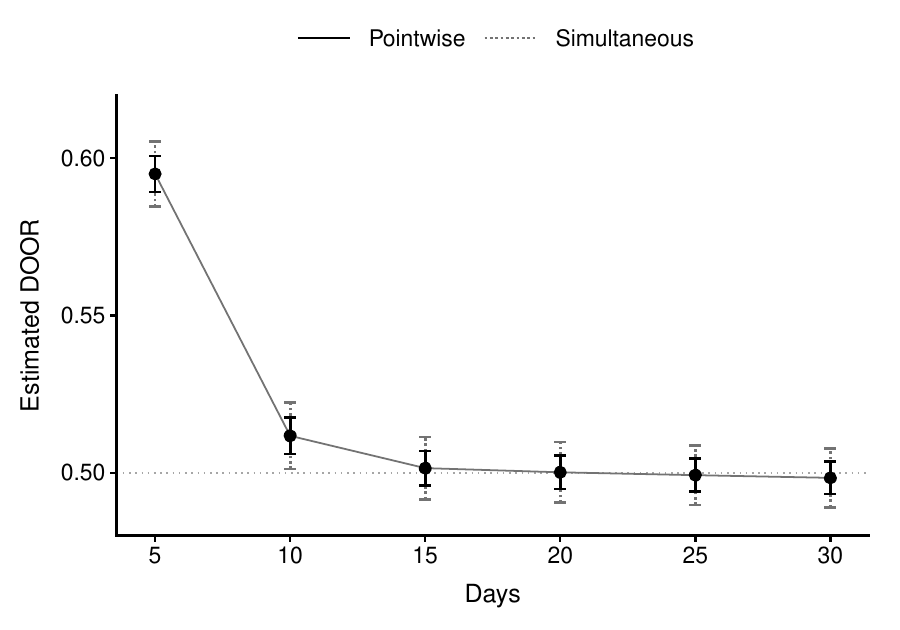}
    \caption{Estimated DOOR trajectory  using TMLE-P with proportional-probit outcome models. }
    \label{fig:application-trajectory}
\par\smallskip
{\small\raggedright\textit{Alt text:} DOOR estimates decline from approximately 0.595 at day 5 to 0.512 at day 10 and remain near 0.50 thereafter. Solid pointwise and dotted simultaneous 95\% intervals exclude 0.50 at days 5 and 10 but include it at later visits.\par}
\end{figure}
\clearpage
\endgroup

\clearpage
\appendix
\renewcommand{\theHsection}{app.\Alph{section}}
\renewcommand{\theHsubsection}{app.\Alph{section}.\arabic{subsection}}
\renewcommand{\theHsubsubsection}{app.\Alph{section}.\arabic{subsection}.\arabic{subsubsection}}

\section*{Appendices}
\newcommand{\appendixsection}[1]{\section{#1}}
\appendixsection{Identification of the target}
\label{app:proof-identification}

\subsection{Proof for Proposition~\ref{prop:identification}}

For treatment \(a\) and visit \(t\), Assumption~\ref{assump::identification} gives
\[
\Prb{Y^a(t)=k}
=\E{\Prb{Y^a(t)=k\mid A=a, X}}
=\E{q_{at}(k\mid X)}
=p_{at,k}.
\]
For independently sampled participants \(i\) and \(j\),
\[
\begin{split}
\psi_t
&=\E{w_{Y_i^1(t)Y_j^0(t)}}\\
&=\sum_{k,\ell}w_{k\ell}
  \Prb{Y_i^1(t)=k}\Prb{Y_j^0(t)=\ell}
=\sum_{k,\ell}w_{k\ell}p_{1t,k}p_{0t,\ell}
=\bm p_{1t}^\top W\bm p_{0t}.
\end{split}
\]
Applying this argument at every visit proves Proposition~\ref{prop:identification}.
\hfill\(\square\)

\subsection{Nonidentification of a within-participant target}
\label{app:within-participant-nonidentification}

Let \(X\) be a constant covariate,
\(A\sim\operatorname{Bernoulli}(1/2)\) be independent of all
potential outcomes, and \(Y(t)=Y^A(t)\). At visit \(t\),
let
\[
(Y^1(t),Y^0(t))\in\{(1,1),(2,2),(3,3)\}
\]
with probability \(1/3\) for each pair in Population 1.
In Population 2, let
\[
(Y^1(t),Y^0(t))\in\{(2,1),(3,2),(1,3)\}
\]
with probability \(1/3\) for each pair. In both populations,
hold each participant's selected pair fixed across visits.

Both populations satisfy Assumption~\ref{assump::identification}
and induce the same observed-data distribution, with \(Y(t)\)
uniform on \(\{1,2,3\}\) in each treatment arm. Nevertheless,
\[
\E{w_{Y^1(t),Y^0(t)}}
=
\begin{cases}
\tfrac12, & \text{in Population 1},\\[2pt]
\tfrac23, & \text{in Population 2}.
\end{cases}
\]
Thus, the within-participant win probability is not identified
under the stated assumptions.

In contrast, the between-participant estimand \(\psi_t\)
equals \(1/2\) in both populations. This estimand compares
treatment-specific marginal outcome distributions and does
not represent the probability that the same participant
would have a more favorable outcome under one treatment than
under the other.

\appendixsection{Doubly-robust Estimation}

\subsection{Efficient influence functions}
\label{app:influence}

For each visit \(t\), the cross-sectional results of
\citet{feng2026}, applied to the marginal record
\(O_t=(X,A,Y(t))\), give the category influence functions
in~\eqref{eq:eifp}. The chain rule applied to
\(\psi_t=\bm p_{1t}^\top W\bm p_{0t}\) then yields
\eqref{eq:eifpsi}. Stacking the visit-specific gradients
gives \(\bm\phi=(\phi_1,\ldots,\phi_T)^\top\) for the
trajectory. Finally, linearity of influence functions gives
\(\phi_\omega=\sum_{t=1}^T\omega_t\phi_t\)
for the prespecified weighted summary.
\hfill\(\square\)

\subsection{Proof of Theorem~\ref{thm:consistency-efficiency}}
\label{app:proof-theorem1}

\subsubsection{Regularity conditions for estimation}
\label{app:est-condition}

\phantomsection
\refstepcounter{assumption}
\paragraph{Assumption \theassumption{} (Estimation conditions)}
\label{assump:estimation}
\begin{enumerate}[label=(\roman*)]
\item \textbf{Conditions for all estimators.}
Under the setup in Section~\ref{sec:targets}, use the
participant-level cross-fitting scheme in Section~\ref{sec:estimation},
with fold assignment independent of the observations.
Let \(K,T,V\) be fixed, with \(V\geq2\) and
\(n_v/n\to1/V\) for every fold.
For every \(a,t,v\), the initial category probabilities satisfy
\[
\widehat q_{at}^{(-v)}(k\mid X)\geq0
\quad(k=1,\ldots,K),
\qquad
\sum_{k=1}^K\widehat q_{at}^{(-v)}(k\mid X)=1
\quad P_X\text{-almost surely}.
\]
Use the clipped propensity predictions and threshold
\(\varepsilon\) defined in Section~\ref{sec:estimation}.

\item \textbf{Additional conditions for TMLE and TMLE-P.}
For every \(a,t,k\), assume \(P_0\{A=a,Y(t)=k\}>0\).
There exists a fixed \(\delta\in(0,1/K)\) such that
\[
\Prb{
\min_{i,a,t,k}
\widehat q_{at}^{(-v(i))}(k\mid X_i)\geq\delta
}
\longrightarrow1.
\]
This lower bound is imposed only at the observed
evaluation points.
\end{enumerate}

\subsubsection{Proof}

Write \(P_0f=\E{f(O)}\),
\(P_nf=n^{-1}\sum_{i=1}^nf(O_i)\), and
\(\lambda_v=n_v/n\).
We first control the targeting updates in TMLE and TMLE-P, then derive a common expansion for the marginal estimators and finally apply
the bilinear DOOR map. All bounds below hold uniformly
over the finitely many arms, visits, categories, and folds.

\paragraph{Step 1: Existence and size of the targeting updates}
Fix an arm \(a\), visit \(t\), and fold \(v\).
We first establish existence, uniqueness, and boundedness
of the targeting minimizers, then derive their rates.

By Assumption~\ref{assump:estimation}(ii) and the law of
large numbers, there exists \(c_0>0\) such that the event
\[
\mathcal A_n
=
\left\{
\min_{v,a,t,k}P_vI\{A=a,Y(t)=k\}\geq c_0,\quad
\min_{i,a,t,k}
\widehat q_{at}^{(-v(i))}(k\mid X_i)\geq\delta
\right\}
\]
has probability tending to one.
On this event, let \(M=\max_{k\leq K}\epsilon_k\) and
\(m=\min_{k\leq K}\epsilon_k\).
For \(i\in I_v\), the denominator in~\eqref{eq:fluct}
is at least
\(\delta\exp\{M/\widehat\pi_a^{(-v)}(X_i)\}\),
while the numerator is at most
\(\exp\{\epsilon_k/\widehat\pi_a^{(-v)}(X_i)\}\).
Hence
\[
-\log\widehat q_{at}^{\bm\epsilon,v}(k\mid X_i)
\geq\log\delta+
\frac{M-\epsilon_k}{\widehat\pi_a^{(-v)}(X_i)}
\geq\log\delta+M-\epsilon_k.
\]
Consequently,
\[
L_{atv}(\bm\epsilon)
\geq P_vI(A=a)\log\delta
+\sum_{k=1}^K P_vI\{A=a,Y(t)=k\}(M-\epsilon_k)
\geq\log\delta+c_0(M-m),
\]
where the second inequality uses \(\log\delta<0\) and
the category attaining \(m\).
At zero, the initial probability floor gives
\(L_{atv}(\bm0)\leq-\log\delta\).
Since \(\epsilon_K=0\),
\(M-m\geq\|\bm\epsilon\|_\infty\), and therefore
\[
L_{atv}(\bm\epsilon)
\geq\log\delta+c_0\|\bm\epsilon\|_\infty
>-\log\delta
\geq L_{atv}(\bm0)
\]
whenever
\(\|\bm\epsilon\|_\infty>2\log(1/\delta)/c_0\).
Thus every minimizer lies in \([-B,B]^{K-1}\) for any
fixed \(B>2\log(1/\delta)/c_0\).

For \(\|\bm\epsilon\|_\infty\leq B\), the numerator
in~\eqref{eq:fluct} is at least
\(\delta e^{-B/\varepsilon}\), while its denominator
is at most \(e^{B/\varepsilon}\).
Hence
\(\widehat q_{at}^{\bm\epsilon,v}(k\mid X_i)
\geq\delta e^{-2B/\varepsilon}\)
for every \(i\in I_v\).
For \(\bm z\in\mathbb R^{K-1}\), set \(z_K=0\).
Differentiating the loss twice gives
\begin{equation}
\bm z^\top\nabla^2L_{atv}(\bm\epsilon)\bm z
=
P_v\left[
\frac{I(A=a)}{\{\widehat\pi_a^{(-v)}(X)\}^2}
\sum_{1\leq j<k\leq K}
\widehat q_{at}^{\bm\epsilon,v}(j\mid X)
\widehat q_{at}^{\bm\epsilon,v}(k\mid X)
(z_j-z_k)^2
\right].
\label{eq:targeting-hessian}
\end{equation}
In~\eqref{eq:targeting-hessian}, retaining only the
pairs \((j,K)\) and using \(P_vI(A=a)\geq Kc_0\) yields
\[
\bm z^\top\nabla^2L_{atv}(\bm\epsilon)\bm z
\geq Kc_0\delta^2e^{-4B/\varepsilon}\|\bm z\|^2
=:\kappa\|\bm z\|^2.
\]
The loss is continuous and exceeds \(L_{atv}(\bm0)\)
outside the fixed cube, so it attains a global minimum
within the cube. The positive curvature bound further ensures
uniqueness. Thus, on \(\mathcal A_n\),
\(\widehat{\bm\epsilon}_{at}^{\,v}\) exists uniquely and
satisfies \(\|\widehat{\bm\epsilon}_{at}^{\,v}\|_\infty\leq B\).
The same bounds apply to the pooled loss
\(L_{atP}(\bm\epsilon)=\sum_v\lambda_vL_{atv}(\bm\epsilon)\),
establishing existence, uniqueness, and the same bound
for \(\widehat{\bm\epsilon}_{at}^{\,\mathrm{TMLE\text{-}P}}\)
on \(\mathcal A_n\).

We next derive the rates of the targeting parameters.
Since
\(\nabla L_{atv}(\widehat{\bm\epsilon}_{at}^{\,v})=\bm0\),
integration along the segment from zero to the minimizer
gives
\[
-\nabla L_{atv}(\bm0)
=
\left\{
\int_0^1
\nabla^2L_{atv}(s\widehat{\bm\epsilon}_{at}^{\,v})\,ds
\right\}
\widehat{\bm\epsilon}_{at}^{\,v}.
\]
Applying the curvature bound and Cauchy--Schwarz to
both losses gives, on \(\mathcal A_n\),
\begin{equation}
\|\widehat{\bm\epsilon}_{at}^{\,v}\|
\leq\kappa^{-1}\|\nabla L_{atv}(\bm0)\|,
\qquad
\|\widehat{\bm\epsilon}_{at}^{\,\mathrm{TMLE\text{-}P}}\|
\leq\kappa^{-1}\|\nabla L_{atP}(\bm0)\|,
\label{eq:targeting-gradient-bound}
\end{equation}
where
\(\nabla L_{atP}(\bm0)
=\sum_v\lambda_v\nabla L_{atv}(\bm0)\).
For \(k<K\), the \(k\)-th component of
the gradient at zero is
\[
\begin{aligned}
[\nabla L_{atv}(\bm0)]_k
&=
P_v\left[
\frac{I(A=a)}{\widehat\pi_a^{(-v)}(X)}
\{\widehat q_{at}^{(-v)}(k\mid X)-I(Y(t)=k)\}
\right]\\
&=
\{P_0+(P_v-P_0)\}\left[
\frac{I(A=a)}{\widehat\pi_a^{(-v)}(X)}
\{\widehat q_{at}^{(-v)}(k\mid X)-I(Y(t)=k)\}
\right].
\end{aligned}
\]
Conditional on the training sample for fold \(v\), the
population term is bounded in absolute value by
\(Cb_{n,v}\), by conditioning on \(X\) and applying
Cauchy--Schwarz. The empirical-process term is
\(O_p(n_v^{-1/2})\) by conditional Chebyshev's inequality
and propensity clipping. Since \(K\) is fixed,
\(\|\nabla L_{atv}(\bm0)\|
=O_p(b_{n,v}+n_v^{-1/2})\).
Summing the fold-specific bounds, using fixed \(V\)
and \(n_v\asymp n\), gives
\(\|\nabla L_{atP}(\bm0)\|=O_p(b_n+n^{-1/2})\).
These conditional bounds are applied separately within
each fold, without conditioning on \(\mathcal A_n\)
or requiring independence among the training samples.
Applying~\eqref{eq:targeting-gradient-bound} and using
\(\Prb{\mathcal A_n}\to1\) yields
\begin{equation}
\|\widehat{\bm\epsilon}_{at}^{\,v}\|
=O_p(b_{n,v}+n_v^{-1/2}),
\qquad
\|\widehat{\bm\epsilon}_{at}^{\,\mathrm{TMLE\text{-}P}}\|
=O_p(b_n+n^{-1/2}).
\label{eq:epsrate}
\end{equation}

Finally, for \(j<K\),
\[
\frac{\partial
\widehat q_{at}^{\bm\epsilon,v}(k\mid x)}
{\partial\epsilon_j}
=
\frac{\widehat q_{at}^{\bm\epsilon,v}(k\mid x)}
{\widehat\pi_a^{(-v)}(x)}
\{I(k=j)-\widehat q_{at}^{\bm\epsilon,v}(j\mid x)\}.
\]
These derivatives are uniformly bounded by
\(1/\varepsilon\) for \(P_X\)-almost every \(x\).
Since \(\widehat q_{at}^{\bm0,v}=\widehat q_{at}^{(-v)}\),
integration along the fluctuation path gives
\[
\max_k
\|\widehat q_{at}^{\bm\epsilon,v}(k\mid\cdot)
-\widehat q_{at}^{(-v)}(k\mid\cdot)\|_2
\leq C\|\bm\epsilon\|.
\]
Combining this inequality with~\eqref{eq:epsrate}
and the triangle inequality yields
\begin{equation}
\begin{aligned}
\max_k
\|\widehat q_{at}^{*,\mathrm{TMLE},v}(k\mid\cdot)
-q_{at}(k\mid\cdot)\|_2
&=O_p(b_{n,v}+n_v^{-1/2}),\\
\max_k
\|\widehat q_{at}^{*,\mathrm{TMLE\text{-}P},v}(k\mid\cdot)
-q_{at}(k\mid\cdot)\|_2
&=O_p(b_{n,v}+b_n+n^{-1/2}).
\end{aligned}
\label{eq:targeted-rates}
\end{equation}
Since \(\Prb{\mathcal A_n}\to1\), these rates hold without conditioning on \(\mathcal A_n\).

\paragraph{Step 2: A common marginal expansion}
For a positive propensity function \(g\) and a category
regression \(h\), define
\[
m_{at,k}(O,g,h)
=
h(k\mid X)
+\frac{I(A=a)}{g(X)}
\{I(Y(t)=k)-h(k\mid X)\}.
\]
At the true nuisance functions,
\(m_{at,k}(O,\pi_a,q_{at})-p_{at,k}
=\varphi_{at,k}(O,P_0)\).

To express all three estimators in common notation, let
\begin{equation}
\widetilde q_{at}^{\,r,v}
=
\begin{cases}
\widehat q_{at}^{(-v)}, & r=\mathrm{AIPW},\\
\widehat q_{at}^{*,r,v},
& r\in\{\mathrm{TMLE},\mathrm{TMLE\text{-}P}\}.
\end{cases}
\label{eq:qtilde}
\end{equation}
The AIPW marginal estimator is the empirical average of
\(m_{at,k}\) evaluated at the initial nuisance estimates.
The targeting equations give the corresponding identity
within each fold for TMLE and after pooling for TMLE-P.
Thus, for all three methods,
\[
\widehat p_{at,k}^{\,r}
=
\sum_v\lambda_vP_vm_{at,k}
(O,\widehat\pi_a^{(-v)},\widetilde q_{at}^{\,r,v}).
\]
Adding and subtracting \(m_{at,k}(O,\pi_a,q_{at})\)
within each fold and splitting \(P_v=P_0+(P_v-P_0)\)
yields the exact decomposition
\begin{equation}
\widehat p_{at,k}^{\,r}-p_{at,k}
=
P_n\varphi_{at,k}(O,P_0)
+\mathcal E_{at,k}^{\,r}
+\mathcal R_{at,k}^{\,r},
\label{eq:marginal-decomposition}
\end{equation}
where
\[
\begin{aligned}
\mathcal E_{at,k}^{\,r}
&=
\sum_v\lambda_v(P_v-P_0)
\left[
m_{at,k}(O,\widehat\pi_a^{(-v)},\widetilde q_{at}^{\,r,v})
-m_{at,k}(O,\pi_a,q_{at})
\right],\\
\mathcal R_{at,k}^{\,r}
&=
\sum_v\lambda_vP_0
\left[
m_{at,k}(O,\widehat\pi_a^{(-v)},\widetilde q_{at}^{\,r,v})
-m_{at,k}(O,\pi_a,q_{at})
\right]\\
&=
\sum_v\lambda_vP_0\left[
\frac{\widehat\pi_a^{(-v)}(X)-\pi_a(X)}
{\widehat\pi_a^{(-v)}(X)}
\{\widetilde q_{at}^{\,r,v}(k\mid X)-q_{at}(k\mid X)\}
\right].
\end{aligned}
\]
The final equality follows by conditioning on \(X\).
Here \(P_0\) integrates a fresh observation while holding
the fitted functions fixed. The terms
\(\mathcal E_{at,k}^{\,r}\) and \(\mathcal R_{at,k}^{\,r}\)
are the empirical-process remainder and the second-order
von Mises remainder, respectively.

\paragraph{Step 3: Bounds for the marginal remainders}
We first bound the empirical-process remainders, then
the second-order von Mises remainders.
Write \(\mathcal E_{at,k}^{\,r,v}\) for the fold-specific
empirical-process term, so that
\(\mathcal E_{at,k}^{\,r}
=\sum_v\lambda_v\mathcal E_{at,k}^{\,r,v}\).

For AIPW, the triangle inequality, positivity, and
propensity clipping imply
\[
\left\|
m_{at,k}(O,\widehat\pi_a^{(-v)},\widehat q_{at}^{(-v)})
-m_{at,k}(O,\pi_a,q_{at})
\right\|_{L_2(P_0)}
\lesssim (a_{n,v}+b_{n,v}).
\]
Conditional on the training sample for fold \(v\),
the fitted functions are fixed and the evaluation
observations are independent. Conditional Chebyshev's
inequality, as in Lemma~1 of \citet{kennedy2024}, gives
\[
\mathcal E_{at,k}^{\,\mathrm{AIPW},v}
=O_p\{(a_{n,v}+b_{n,v})n_v^{-1/2}\}.
\]
Summing over fixed \(V\), with \(n_v\asymp n\), yields
$\mathcal E_{at,k}^{\,\mathrm{AIPW}} =O_p\{(a_n+b_n)n^{-1/2}\}$.

For \(r\in\{\mathrm{TMLE},\mathrm{TMLE\text{-}P}\}\),
the additional empirical-process term within fold \(v\) is
\[
\mathcal E_{at,k}^{\,r,v}
-\mathcal E_{at,k}^{\,\mathrm{AIPW},v}
=
(P_v-P_0)\left[
\left\{1-\frac{I(A=a)}{\widehat\pi_a^{(-v)}(X)}\right\}
\{\widehat q_{at}^{*,r,v}(k\mid X)
-\widehat q_{at}^{(-v)}(k\mid X)\}
\right].
\]
The targeted outcome model is fitted using the evaluation
outcomes, so the preceding conditional variance argument
does not apply directly. Instead, define
\[
H_{atkv}(O,\bm\epsilon)
=
\left\{1-\frac{I(A=a)}{\widehat\pi_a^{(-v)}(X)}\right\}
\widehat q_{at}^{\bm\epsilon,v}(k\mid X).
\]
On the event
\(\|\widehat{\bm\epsilon}_{at}^{\,r,v}\|_\infty\leq B\),
the fundamental theorem of calculus and Cauchy--Schwarz
give
\[
\left|
\mathcal E_{at,k}^{\,r,v}
-\mathcal E_{at,k}^{\,\mathrm{AIPW},v}
\right|
\leq
\|\widehat{\bm\epsilon}_{at}^{\,r,v}\|
\sup_{\|\bm\epsilon\|_\infty\leq B}
\left\|(P_v-P_0)\nabla_{\bm\epsilon}
H_{atkv}(O,\bm\epsilon)\right\|.
\]
Conditional on the training sample, the coordinate
derivatives of \(H_{atkv}(O,\bm\epsilon)\) form function
classes indexed by \(\bm\epsilon\in[-B,B]^{K-1}\).
They are uniformly bounded and Lipschitz in
\(\bm\epsilon\), with constants independent of the
training sample. 
The class therefore has a uniformly bounded entropy
integral. A conditional empirical-process maximal
inequality followed by Markov's inequality gives
\[
\sup_{\|\bm\epsilon\|_\infty\leq B}
\left\|(P_v-P_0)\nabla_{\bm\epsilon}
H_{atkv}(O,\bm\epsilon)\right\|
=O_p(n_v^{-1/2}),
\]
using the conditional empirical-process approach of
\citet{zheng2010asymptotic}. 

Combining this bound with~\eqref{eq:epsrate} and the
AIPW bound gives the fold-specific orders
\[
\begin{aligned}
\mathcal E_{at,k}^{\,\mathrm{TMLE},v}
&=O_p\{(a_{n,v}+b_{n,v})n_v^{-1/2}+n_v^{-1}\},\\
\mathcal E_{at,k}^{\,\mathrm{TMLE\text{-}P},v}
&=O_p\{(a_{n,v}+b_{n,v}+b_n)n_v^{-1/2}
       +n^{-1/2}n_v^{-1/2}\}.
\end{aligned}
\]
For TMLE-P, the supremum is taken before substituting
the shared fluctuation, so the bound remains valid even
though that fluctuation depends on all folds.
Summing over folds yields, for either targeting scheme,
\[
\mathcal E_{at,k}^{\,r}
=O_p\{(a_n+b_n)n^{-1/2}+n^{-1}\}.
\]

For the von Mises remainder, propensity clipping and
Cauchy--Schwarz give
\[
|\mathcal R_{at,k}^{\,r}|
\leq C\sum_v\lambda_v a_{n,v}
\|\widetilde q_{at}^{\,r,v}(k\mid\cdot)
-q_{at}(k\mid\cdot)\|_2.
\]
Using the initial outcome errors for AIPW and
\eqref{eq:targeted-rates} for the targeting schemes,
we obtain
\[
|\mathcal R_{at,k}^{\,\mathrm{AIPW}}|
\leq C\sum_v\lambda_v a_{n,v}b_{n,v},
\]
\[
\mathcal R_{at,k}^{\,\mathrm{TMLE}}
=O_p\left\{\sum_v\lambda_v a_{n,v}b_{n,v}
+a_n n^{-1/2}\right\}, \quad 
\mathcal R_{at,k}^{\,\mathrm{TMLE\text{-}P}}
=O_p\{a_nb_n+a_n n^{-1/2}\},
\]
where we used $\max_v b_{n,v} \leq {b_n}/{\min_v\lambda_v}
=O_p(b_n)$ in the bound for TMLE-P.

Combining these bounds with~\eqref{eq:marginal-decomposition}
gives, for \(r\in\{\mathrm{AIPW},\mathrm{TMLE}\}\),
\[
\widehat p_{at,k}^{\,r}-p_{at,k}
=
P_n\varphi_{at,k}(O,P_0)
+O_p\left\{(a_n+b_n)n^{-1/2}+n^{-1}
+\sum_v\lambda_v a_{n,v}b_{n,v}\right\},
\]
whereas
\[
\widehat p_{at,k}^{\,\mathrm{TMLE\text{-}P}}-p_{at,k}
=
P_n\varphi_{at,k}(O,P_0)
+O_p\{(a_n+b_n)n^{-1/2}+n^{-1}+a_nb_n\}.
\]
The initial nuisance errors are bounded, and
\(P_n\varphi_{at,k}(O,P_0)=O_p(n^{-1/2})\).
The consistency conditions therefore imply
\(\widehat{\bm p}_{at}^{\,r}\xrightarrow{p}\bm p_{at}\).
Under the efficiency conditions, the remainders are
\(o_p(n^{-1/2})\), giving
\begin{equation}
\widehat{\bm p}_{at}^{\,r}-\bm p_{at}
=
P_n\bm{\varphi}_{at}(O,P_0)+o_p(n^{-1/2})
\label{eq:pooled-marginal-expansion}
\end{equation}
for all three methods. 
For AIPW and fold-specific TMLE, the same argument also
gives a root-\(n\) expansion for each fold-specific marginal
estimator. For TMLE-P, the expansion is established only
for the pooled marginal estimator and does not require
a root-\(n\) expansion within each fold.

\paragraph{Step 4: The DOOR trajectory and weighted summaries}
For each method \(r\), let
\(\bm\Delta_{at}^{\,r}
=\widehat{\bm p}_{at}^{\,r}-\bm p_{at}\).
By the bilinear representation in~\eqref{eq:id},
\[
\widehat\psi_t^{\,r}-\psi_t
=
(\bm\Delta_{1t}^{\,r})^\top W\bm p_{0t}
+\bm p_{1t}^\top W\bm\Delta_{0t}^{\,r}
+(\bm\Delta_{1t}^{\,r})^\top W\bm\Delta_{0t}^{\,r}.
\]
Under the consistency conditions, the pooled marginal
estimators are consistent, so every term on the right
is \(o_p(1)\). Since \(T\) is fixed, consistency follows
for both the trajectory and its prespecified weighted
summaries.

Under the efficiency conditions,
\eqref{eq:pooled-marginal-expansion} gives
\(\bm\Delta_{at}^{\,r}=O_p(n^{-1/2})\), making the
quadratic term \(O_p(n^{-1})\).
Substituting the marginal expansions into the linear
terms and using~\eqref{eq:eifpsi} yields
\[
\widehat\psi_t^{\,r}-\psi_t
=
P_n\phi_t(O,P_0)+o_p(n^{-1/2}).
\]
Stacking over visits and taking the prespecified
weighted sum gives
\[
\widehat{\bm\psi}^{\,r}-\bm\psi
=P_n\bm\phi(O,P_0)+o_p(n^{-1/2}),
\qquad
\widehat\Psi_\omega^{\,r}-\Psi_\omega
=P_n\phi_\omega(O,P_0)+o_p(n^{-1/2}).
\]
Thus, the corresponding asymptotic variances equal the
semiparametric efficiency bounds, completing the proof.
\hfill\(\square\)

\subsection{Direct one-step estimator}
\label{app:direct-estimator}

A direct one-step estimator corrects the DOOR plug-in
estimator using its influence function, whereas the
proposed AIPW construction corrects the marginal category
probabilities before forming DOOR.
To compare these approaches, define
\[
\widehat{\bm p}_{at}^{\,g,v}
=P_v\widehat{\bm q}_{at}^{(-v)}(X),
\qquad
\widehat{\bm p}_{at}^{\,g}
=\sum_v\lambda_v\widehat{\bm p}_{at}^{\,g,v},
\qquad
\widehat\psi_t^{\,g}
=(\widehat{\bm p}_{1t}^{\,g})^\top
W\widehat{\bm p}_{0t}^{\,g}.
\]
Let \(\bm{\varphi}_{at}(O,\widehat P_v)\) denote the
category influence functions evaluated at the initial
nuisance fits for fold \(v\), centered at
\(\widehat{\bm p}_{at}^{\,g,v}\).
Using the pooled marginal vectors as the coefficients
in~\eqref{eq:eifpsi}, define the direct one-step estimator as
\[
\widehat\psi_t^{\,\mathrm{OS}}
=\widehat\psi_t^{\,g}
+\left\{\sum_v\lambda_vP_v
\bm{\varphi}_{1t}(O,\widehat P_v)\right\}^{\top}
W\widehat{\bm p}_{0t}^{\,g}
+(\widehat{\bm p}_{1t}^{\,g})^\top W
\left\{\sum_v\lambda_vP_v
\bm{\varphi}_{0t}(O,\widehat P_v)\right\}.
\]
The correction is thus applied directly to
\(\widehat\psi_t^{\,g}\).

In contrast, our AIPW construction first forms
\(\widehat{\bm p}_{at}^{\,\mathrm{AIPW}}
=\widehat{\bm p}_{at}^{\,g}
+\sum_v\lambda_vP_v\bm{\varphi}_{at}(O,\widehat P_v)\),
then combines the corrected marginals through
\eqref{eq:aipwpsi}.
Expanding this product gives the exact identity
\begin{equation}
\widehat\psi_t^{\,\mathrm{AIPW}}
-\widehat\psi_t^{\,\mathrm{OS}}
=
\left\{\sum_v\lambda_vP_v
\bm{\varphi}_{1t}(O,\widehat P_v)\right\}^{\top}
W
\left\{\sum_v\lambda_vP_v
\bm{\varphi}_{0t}(O,\widehat P_v)\right\}.
\label{eq:direct-aipw-difference}
\end{equation}
Under the AIPW efficiency conditions in
Theorem~\ref{thm:consistency-efficiency}, the direct
estimator shares the efficient expansion if and only
if this product is \(o_p(n^{-1/2})\) at each visit.

To obtain a sufficient condition, apply the marginal
expansion in~\eqref{eq:pooled-marginal-expansion}:
\[
\begin{aligned}
\sum_v\lambda_vP_v\bm{\varphi}_{at}(O,\widehat P_v)
&=\widehat{\bm p}_{at}^{\,\mathrm{AIPW}}
-\widehat{\bm p}_{at}^{\,g}
=\bm p_{at}-\widehat{\bm p}_{at}^{\,g}
+O_p(n^{-1/2})\\
&=-\sum_v\lambda_vP_0
\{\widehat{\bm q}_{at}^{(-v)}(X)-\bm q_{at}(X)\}
\\
&\quad-\sum_v\lambda_v(P_v-P_0)\widehat{\bm q}_{at}^{(-v)}(X)
+O_p(n^{-1/2})\\
&=-\sum_v\lambda_vP_0
\{\widehat{\bm q}_{at}^{(-v)}(X)-\bm q_{at}(X)\}
+O_p(n^{-1/2}),
\end{aligned}
\]
where the final equality follows from conditional
cross-fitting and bounded category probabilities.
Outcome consistency makes the pooled population errors
\(o_p(1)\). Substitution into~\eqref{eq:direct-aipw-difference}
therefore yields
\[
\begin{aligned}
&\widehat\psi_t^{\,\mathrm{AIPW}}
-\widehat\psi_t^{\,\mathrm{OS}}\\
&={}
\left[
\sum_v\lambda_vP_0
\{\widehat{\bm q}_{1t}^{(-v)}(X)-\bm q_{1t}(X)\}
\right]^\top
W
\left[
\sum_v\lambda_vP_0
\{\widehat{\bm q}_{0t}^{(-v)}(X)-\bm q_{0t}(X)\}
\right]
+o_p(n^{-1/2}),
\end{aligned}
\]
because the cross terms are \(o_p(n^{-1/2})\) and the
product of the stochastic remainders is \(O_p(n^{-1})\).
Since
\(|\bm x^\top W\bm y|
\leq\|W\|_{\mathrm{op}}\|\bm x\|\,\|\bm y\|\),
a sufficient additional condition is
\[
\left\|
\sum_v\lambda_vP_0
\{\widehat{\bm q}_{1t}^{(-v)}(X)-\bm q_{1t}(X)\}
\right\|
\left\|
\sum_v\lambda_vP_0
\{\widehat{\bm q}_{0t}^{(-v)}(X)-\bm q_{0t}(X)\}
\right\|
=o_p(n^{-1/2}).
\]
This outcome-only condition is not generally implied
by the AIPW efficiency conditions, which control
products of propensity and outcome regression errors.

\appendixsection{Inference under nuisance-model misspecification}
\label{app:inference}

Retain \(P_0,P_n,P_v\), and \(\lambda_v\) from the previous
proof such that \(P_nf=\sum_v\lambda_vP_vf\), and define the training-sample average by
\[
P_{-v}f=\frac1{n-n_v}\sum_{i\notin I_v}f(O_i).
\]
The sampling and fold conditions are given in
Assumption~\ref{assump:estimation}.

For a parametric propensity model, write
\(\pi_{\mathrm{raw}}(x;\bm\beta)\) for the raw prediction and set
\[
\pi(x;\bm\beta)
=\min\{1-\varepsilon,\max\{\varepsilon,
\pi_{\mathrm{raw}}(x;\bm\beta)\}\},
\qquad
\pi_a(x;\bm\beta)
=\pi(x;\bm\beta)^a\{1-\pi(x;\bm\beta)\}^{1-a}.
\]
The individual likelihood score uses the raw probability:
\[
\bm s_\beta(O;\bm\beta)
=\frac{\partial}{\partial\bm\beta}
\left[A\log\pi_{\mathrm{raw}}(X;\bm\beta)
+(1-A)\log\{1-\pi_{\mathrm{raw}}(X;\bm\beta)\}\right].
\]
Under correct specification, evaluating at \(\bm\beta_0\)
gives the score and information in~\eqref{eq:propensity-score-information}.
For the logistic model,
\(\pi_{\mathrm{raw}}(x;\bm\beta)
=\operatorname{expit}\{\bm x_\pi(x)^\top\bm\beta\}\) and
\(\bm s_\beta(O;\bm\beta)
=\bm x_\pi(X)\{A-\pi_{\mathrm{raw}}(X;\bm\beta)\}\).

Let
\(\bm z_{it}=(I\{Y_i(t)=1\},\ldots,I\{Y_i(t)=K\})^\top\).
For a category-probability vector \(\bm q(x)\) and
a positive arm-specific propensity \(g(x)\), the
multinomial fluctuation has coordinates
\[
q^{\bm\epsilon}(k\mid x)
=\frac{q(k\mid x)\exp\{\epsilon_k/g(x)\}}
{\sum_{\ell=1}^Kq(\ell\mid x)
\exp\{\epsilon_\ell/g(x)\}},
\qquad \epsilon_K=0.
\]
We write
\(\bm q^{\bm\epsilon}(x)
=(q^{\bm\epsilon}(1\mid x),\ldots,
q^{\bm\epsilon}(K\mid x))^\top\).
For a vector \(\bm u\), let \(\bm u_{[1:m]}\) denote
its first \(m\) coordinates.
We use \(\|\cdot\|_2\) for the \(L_2(P_X)\) norm
of functions and \(\|\cdot\|\) for the Euclidean
norm of vectors.

\subsection{Proof for Proposition~\ref{prop:corrected-parametric}}

We first define the participant-level estimating
contributions, state the regularity conditions, and
then prove Proposition~\ref{prop:corrected-parametric}.

\subsubsection{Estimating contributions}
\label{app:inference-definitions}

Let \(\bm\eta_v\) collect the initial nuisance parameters
estimated using participants outside \(I_v\), including
the propensity parameter \(\bm\beta_v\) and outcome
parameters \(\bm\theta_{at,v}\).
Let \(\bm\gamma\) stack these parameters across folds,
together with the fold-specific fluctuation parameters
for TMLE or the shared fluctuation parameters for TMLE-P.
For simplicity, we omit the method superscript \(r\)
from \(\bm\gamma\) and \(\bm s_i\).

\paragraph{Score contributions}
For AIPW, the estimating equations contain only the
initial nuisance scores. For TMLE and TMLE-P, they
also contain the targeting scores.
Define the working log-likelihood contribution
\[
\begin{aligned}
\ell_\eta(O_i;\bm\eta)
&=A_i\log\pi_{\mathrm{raw}}(X_i;\bm\beta)
+(1-A_i)\log\{1-\pi_{\mathrm{raw}}(X_i;\bm\beta)\}
\\
&\quad+\sum_{a,t}I(A_i=a)
\log q_{at}\{Y_i(t)\mid X_i;\bm\theta_{at}\}.
\end{aligned}
\]
Accordingly, define the normalized training-score block as
\[
\bm s_{\eta_v,i}^{\mathrm{tr}}(\bm\eta_v)
=\frac{n}{n-n_v}I(i\notin I_v)
\frac{\partial\ell_\eta(O_i;\bm\eta_v)}
{\partial\bm\eta_v},
\]
and its average over all participants equals the
training-sample average score.
If the working nuisance model is fixed, its corresponding parameter and score block can be omitted.

Let \(\bm q_{at}(x;\bm\eta_v)\) denote the initial
category-probability vector used in estimation, and write
\(\bm q_{at}^{\bm\epsilon}(x;\bm\eta_v)\)
for its multinomial fluctuation.
For TMLE, additionaly append one targeting block for each \((a,t,v)\):
\[
\bm s_{\epsilon_{at}^{v},i}^{\mathrm{TMLE}}(\bm\gamma)
=\frac{n}{n_v}I(i\in I_v)
\frac{I(A_i=a)}{\pi_a(X_i;\bm\beta_v)}
\left\{\bm z_{it}
-\bm q_{at}^{\bm\epsilon_{at}^{v}}(X_i;\bm\eta_v)
\right\}_{[1:K-1]}.
\]
The factor \(n/n_v\) ensures that averaging over all
participants recovers the targeting score equation
within fold \(I_v\).
For TMLE-P, append one shared targeting block for
each \((a,t)\):
\[
\bm s_{\epsilon_{at},i}^{\mathrm{TMLE\text{-}P}}(\bm\gamma)
=\sum_{v=1}^V I(i\in I_v)
\frac{I(A_i=a)}{\pi_a(X_i;\bm\beta_v)}
\left\{\bm z_{it}
-\bm q_{at}^{\bm\epsilon_{at}^{\mathrm{TMLE\text{-}P}}}
(X_i;\bm\eta_v)\right\}_{[1:K-1]}.
\]
Averaging over all participants recovers the pooled
targeting score equation, with each fold weighted
by \(n_v/n\).
Stacking the training blocks only defines \(\bm s_i(\bm\gamma)\)
for AIPW. Adding the fold-specific or shared targeting
blocks gives the corresponding score vector for
TMLE or TMLE-P.

\paragraph{Marginal contributions}
For each arm and visit, define
\[
\bm f_{at,i}^{\,r}(\bm\gamma)
=
\begin{cases}
\displaystyle
\bm q_{at}(X_i;\bm\eta_{v(i)})
+\frac{I(A_i=a)}{\pi_a(X_i;\bm\beta_{v(i)})}
\{\bm z_{it}-\bm q_{at}(X_i;\bm\eta_{v(i)})\},
& r=\mathrm{AIPW},\\[6pt]
\bm q_{at}^{\bm\epsilon_{at}^{v(i)}}
(X_i;\bm\eta_{v(i)}),
& r=\mathrm{TMLE},\\[3pt]
\bm q_{at}^{\bm\epsilon_{at}^{\mathrm{TMLE\text{-}P}}}
(X_i;\bm\eta_{v(i)}),
& r=\mathrm{TMLE\text{-}P}.
\end{cases}
\]
Evaluating these contributions at the fitted parameters
and averaging over participants gives
\[
\widehat{\bm p}_{at}^{\,r}
=\frac1n\sum_{i=1}^n
\bm f_{at,i}^{\,r}(\widehat{\bm\gamma}).
\]

\subsubsection{Regularity conditions}
\label{app:inference-conditions}

The following conditions are used to prove
Proposition~\ref{prop:corrected-parametric}.
\phantomsection
\refstepcounter{assumption}
\paragraph{Assumption \theassumption{} (Parametric inference)}
\label{assump:inference}
Assume Assumptions~\ref{assump::identification}
and~\ref{assump:estimation}.
Let \(\widehat{\bm\gamma}\) solve
\[
n^{-1}\sum_{i=1}^n\bm s_i(\widehat{\bm\gamma})=\bm0.
\]
Suppose the following additional conditions hold.
\begin{enumerate}[label=(\roman*)]
\item \textbf{Consistency.}
The parameter dimension is fixed, and
\(\widehat{\bm\gamma}\xrightarrow{p}\bm\gamma^\star\)
for a finite deterministic interior limit.
Initial nuisance limits are common across folds.
For TMLE and TMLE-P, the limiting initial category
probabilities are positive almost surely, and the
population targeting loss has a finite minimizer.

\item \textbf{Smoothness and moments.}
On a neighborhood \(\mathcal N\) of \(\bm\gamma^\star\),
the score and marginal contributions are continuously
differentiable in \(\bm\gamma\) almost surely, with
\[
P_0\|\bm s_i(\bm\gamma^\star)\|^2<\infty,
\qquad
P_0\|\bm f_{at,i}^{\,r}(\bm\gamma^\star)\|^2<\infty.
\]
There exists an envelope \(M(O)\), independent of the
fold and sample size, with \(P_0M^2<\infty\), such that
\[
\sup_{\bm\gamma\in\mathcal N}
\left\{
\left\|\frac{\partial\bm s_i(\bm\gamma)}
{\partial\bm\gamma^\top}\right\|_{\mathrm{op}}
+\max_{a,t}
\left\|\frac{\partial\bm f_{at,i}^{\,r}(\bm\gamma)}
{\partial\bm\gamma^\top}\right\|_{\mathrm{op}}
\right\}
\leq M(O_i)
\]
for all participants and sufficiently large sample sizes.
The matrix \(\bm A\) is nonsingular.
\end{enumerate}

\subsubsection{Proof}
\label{app:proof-prop2}

\paragraph{Step 1: Parametric and marginal expansions}
Consistency and the uniform law of large numbers imply
that the limiting nuisance parameters and fluctuation
parameters solve their respective population estimating
equations.
For TMLE and TMLE-P, positive limiting category
probabilities make the population targeting loss
strictly convex, so its finite minimizer is unique.
Common initial nuisance limits therefore give common
targeting limits across folds.
Consequently,
\[
P_0\bm s_i(\bm\gamma^\star)=\bm0,
\qquad
P_0\bm f_{at,i}^{\,r}(\bm\gamma^\star)
=\bm p_{at}^{\,r,\star},
\]
where expectations condition on the fold assignment.
At \(\bm\gamma^\star\), the contributions are independent
across participants conditional on this assignment.
Within each fold, they are functions of identically
distributed records, with deterministic normalizing
factors converging to finite limits.
Finite second moments and fixed \(V\) therefore yield
a joint central limit theorem for the centered score
and marginal averages.

To expand the estimating equations, define
\(\bm\gamma_u
=\bm\gamma^\star+u(\widehat{\bm\gamma}-\bm\gamma^\star)\),
\(u\in[0,1]\), and
\[
\overline{\bm A}_n
=\int_0^1\frac1n\sum_{i=1}^n
\frac{\partial\bm s_i(\bm\gamma_u)}
{\partial\bm\gamma^\top}\,du.
\]
By consistency, this segment lies in the smoothness
neighborhood with probability tending to one.
Uniform convergence of the empirical Jacobians and
continuity of their population limits imply
\(\overline{\bm A}_n=\bm A+o_p(1)\).
The fundamental theorem of calculus then gives
\[
\bm0
=\frac1n\sum_{i=1}^n\bm s_i(\bm\gamma^\star)
+\overline{\bm A}_n
(\widehat{\bm\gamma}-\bm\gamma^\star).
\]
The centered score average is \(O_p(n^{-1/2})\)
by a conditional variance bound.
Since \(\bm A\) is nonsingular,
\(\overline{\bm A}_n\) is invertible with probability
tending to one and
\(\overline{\bm A}_n^{-1}=\bm A^{-1}+o_p(1)\).
Hence
\[
\widehat{\bm\gamma}-\bm\gamma^\star
=-\bm A^{-1}\frac1n\sum_{i=1}^n
\bm s_i(\bm\gamma^\star)+o_p(n^{-1/2}).
\]

For the marginal estimator, similarly define
\[
\overline{\bm G}_{at,n}
=\int_0^1\frac1n\sum_{i=1}^n
\frac{\partial\bm f_{at,i}^{\,r}(\bm\gamma_u)}
{\partial\bm\gamma^\top}\,du.
\]
The same argument gives
\(\overline{\bm G}_{at,n}=\bm G_{at}+o_p(1)\).
Since
\(\widehat{\bm\gamma}-\bm\gamma^\star=O_p(n^{-1/2})\),
\[
\widehat{\bm p}_{at}^{\,r}-\bm p_{at}^{\,r,\star}
=\frac1n\sum_{i=1}^n
\{\bm f_{at,i}^{\,r}(\bm\gamma^\star)
-\bm p_{at}^{\,r,\star}\}
+\bm G_{at}(\widehat{\bm\gamma}-\bm\gamma^\star)
+o_p(n^{-1/2}).
\]
Substituting the parameter expansion yields
\[
\widehat{\bm p}_{at}^{\,r}-\bm p_{at}^{\,r,\star}
=\frac1n\sum_{i=1}^n
\bm{\varphi}_{at,i}^{\,r,\mathrm{corr}}
+o_p(n^{-1/2}),
\]
where
\[
\bm{\varphi}_{at,i}^{\,r,\mathrm{corr}}
=\bm f_{at,i}^{\,r}(\bm\gamma^\star)
-\bm p_{at}^{\,r,\star}
-\bm G_{at}\bm A^{-1}\bm s_i(\bm\gamma^\star).
\]

\paragraph{Step 2: DOOR expansion and asymptotic normality}
Write
\(\bm\Delta_{at}^{\,r}
=\widehat{\bm p}_{at}^{\,r}-\bm p_{at}^{\,r,\star}\).
The bilinear DOOR formula gives
\[
\widehat\psi_t^{\,r}-\psi_t^{\,r,\star}
=(\bm\Delta_{1t}^{\,r})^\top W\bm p_{0t}^{\,r,\star}
+(\bm p_{1t}^{\,r,\star})^\top W\bm\Delta_{0t}^{\,r}
+(\bm\Delta_{1t}^{\,r})^\top W\bm\Delta_{0t}^{\,r}.
\]
Since \(\bm\Delta_{at}^{\,r}=O_p(n^{-1/2})\),
the final term is \(O_p(n^{-1})\).
Substituting the marginal expansions into the linear
terms gives
\[
\widehat\psi_t^{\,r}-\psi_t^{\,r,\star}
=\frac1n\sum_{i=1}^n\phi_{t,i}^{\,r,\mathrm{corr}}
+o_p(n^{-1/2}),
\]
where
\[
\phi_{t,i}^{\,r,\mathrm{corr}}
=(\bm{\varphi}_{1t,i}^{\,r,\mathrm{corr}})^\top
W\bm p_{0t}^{\,r,\star}
+(\bm p_{1t}^{\,r,\star})^\top
W\bm{\varphi}_{0t,i}^{\,r,\mathrm{corr}}.
\]
Stacking over visits and applying the prespecified
weights gives the expansions in~\eqref{eq:stackedcorr}.
The corrected contributions are fixed linear combinations
of the centered score and marginal contributions.
Their joint central limit theorem therefore gives
\[
\frac1{\sqrt n}\sum_{i=1}^n
\bm\phi_i^{\,r,\mathrm{corr}}
\xrightarrow{d}N_T(\bm0,\bm\Sigma_r),
\qquad
\bm\Sigma_r
=\lim_{n\to\infty}\frac1n\sum_{i=1}^n
P_0\{\bm\phi_i^{\,r,\mathrm{corr}}
(\bm\phi_i^{\,r,\mathrm{corr}})^\top\}.
\]
Combining this with~\eqref{eq:stackedcorr} and
Slutsky's theorem yields
\[
\sqrt n(\widehat{\bm\psi}^{\,r}-\bm\psi^{\,r,\star})
\xrightarrow{d}N_T(\bm0,\bm\Sigma_r),
\qquad
\sqrt n(\widehat\Psi_\omega^{\,r}-\Psi_\omega^{\,r,\star})
\xrightarrow{d}N(0,\bm\omega^\top\bm\Sigma_r\bm\omega).
\]

\paragraph{Step 3: Centering at the target estimand}
Under the identification conditions, we now show that
the limiting marginal probabilities
\(\bm p_{at}^{\,r,\star}\) equal \(\bm p_{at}\)
when either the propensity predictions or all initial
outcome predictions are consistent.

For AIPW, the difference between the limiting marginal
mean and \(p_{at,k}\) is determined by the population
remainder \(\mathcal R_{at,k}^{\,\mathrm{AIPW}}\)
in~\eqref{eq:marginal-decomposition}.
Bounded category predictions and propensity clipping
ensure that this remainder converges to zero when
either nuisance prediction is consistent, giving
\(p_{at,k}^{\,\mathrm{AIPW},\star}=p_{at,k}\).

For TMLE and TMLE-P, let \(\bm\beta^\star\) denote
the common limit of the fitted propensity parameters
and \(q_{at}^{r,\star}\) the limiting targeted outcome
prediction.
The population targeting equations give
\[
P_0\left[
\frac{I(A=a)}{\pi_a(X;\bm\beta^\star)}
\{I(Y(t)=k)-q_{at}^{r,\star}(k\mid X)\}
\right]=0
\]
for every category, with the final equation following
because the category residuals sum to zero.
Hence the limiting targeted marginal mean equals
its augmented counterpart:
\[
p_{at,k}^{\,r,\star}
=P_0q_{at}^{r,\star}(k\mid X)
=P_0\left[
q_{at}^{r,\star}(k\mid X)
+\frac{I(A=a)}{\pi_a(X;\bm\beta^\star)}
\{I(Y(t)=k)-q_{at}^{r,\star}(k\mid X)\}
\right].
\]
If \(\pi_a(X;\bm\beta^\star)=\pi_a(X)\) almost surely,
conditioning on \(X\) gives
\[
p_{at,k}^{\,r,\star}
=P_0\{q_{at}^{r,\star}(k\mid X)
+q_{at}(k\mid X)-q_{at}^{r,\star}(k\mid X)\}
=p_{at,k}.
\]
If instead the limiting initial outcome prediction
equals \(q_{at}\), zero fluctuation satisfies the
population targeting equations, and  uniqueness of the population minimizer therefore implies that the limiting fluctuation is zero.
Hence \(q_{at}^{r,\star}=q_{at}\), again giving
\(p_{at,k}^{\,r,\star}=p_{at,k}\).

Thus either case yields
\(\bm p_{at}^{\,r,\star}=\bm p_{at}\),
\(\bm\psi^{\,r,\star}=\bm\psi\), and
\(\Psi_\omega^{\,r,\star}=\Psi_\omega\).
\paragraph{Step 4: Covariance consistency}
The smoothness and envelope conditions imply that
the empirical Jacobians evaluated at
\(\widehat{\bm\gamma}\) consistently estimate
\(\bm A\) and \(\bm G_{at}\).
By parameter consistency, the derivative bound applies
along the segment between \(\bm\gamma^\star\) and
\(\widehat{\bm\gamma}\) with probability tending to one,
yielding
\[
\frac1n\sum_{i=1}^n
\|\bm s_i(\widehat{\bm\gamma})
-\bm s_i(\bm\gamma^\star)\|^2
\leq
\|\widehat{\bm\gamma}-\bm\gamma^\star\|^2
\frac1n\sum_{i=1}^nM(O_i)^2=o_p(1),
\]
with the same bound for the marginal contributions.
Together with consistent derivative and marginal
estimates and finite second moments, this implies
\[
\frac1n\sum_{i=1}^n
\|\widehat{\bm\phi}_i^{\,r,\mathrm{corr}}
-\bm\phi_i^{\,r,\mathrm{corr}}\|^2=o_p(1).
\]

The law of large numbers within each fold and
convergence of the fold weights give convergence
of the empirical covariance of the limiting
contributions to \(\bm\Sigma_r\).
By Cauchy--Schwarz, replacing them by their fitted
versions changes this covariance by \(o_p(1)\).
Consequently, the estimator in~\eqref{eq:corrected-covariance} satisfies
\[
n\widehat V_r\xrightarrow{p}\bm\Sigma_r.
\]
For the weighted summary,
\(n\bm\omega^\top\widehat V_r\bm\omega
\xrightarrow{p}\bm\omega^\top\bm\Sigma_r\bm\omega\).
This establishes consistency of the covariance
and variance estimators.
\hfill\(\square\)
\subsection{Proof for Proposition~\ref{prop:correct-ps}}

We first state the regularity conditions, then derive
the corrected expansion and variance identity.
Finally, we describe estimation of the correction
and establish covariance consistency.

\subsubsection{Regularity conditions}
\label{app:projection-conditions}

Use the propensity model, clipped predictions, and likelihood score
defined at the beginning of Appendix~\ref{app:inference}.
Under correct specification, \(\bm s_\beta(O)\) and
\(\bm I_\beta\) are as in~\eqref{eq:propensity-score-information}.

\phantomsection
\refstepcounter{assumption}
\paragraph{Assumption \theassumption{} (Correct propensity model)}
\label{assump:correct-ps}
Assume Assumptions~\ref{assump::identification}
and~\ref{assump:estimation}.
Suppose the following additional conditions hold.
\begin{enumerate}[label=(\roman*)]
\item \textbf{Propensity estimation.}
The propensity model is finite-dimensional and correctly specified such that
\(\pi_{\mathrm{raw}}(X;\bm\beta_0)=\pi(X)\),
where \(\bm\beta_0\) is an interior parameter.
The training-sample MLEs exist with
probability tending to one, solve their likelihood
score equations, and satisfy
\[
\max_{v}
\|\widehat{\bm\beta}^{(-v)}-\bm\beta_0\|=o_p(1).
\]
Near \(\bm\beta_0\), the raw propensity and likelihood
score are continuously differentiable almost surely,
with derivative norms bounded by square-integrable
envelopes.
Assume \(P_0\|\bm s_\beta(O)\|^2<\infty\),
\(\bm I_\beta\) is nonsingular, and \(0<\varepsilon<c\),
where \(c\) is the positivity bound in
Assumption~\ref{assump::identification}.

\item \textbf{Convergence of initial outcome predictions.}
The initial outcome predictions converge to deterministic
limits common across folds:
\[
\max_{a,t,k,v}
\|\widehat q_{at}^{(-v)}(k\mid\cdot)
-q_{at}^{\mathrm{init},\star}(k\mid\cdot)\|_2=o_p(1).
\]
These limits need not equal the true outcome regressions.

\item \textbf{Targeting.}
For TMLE and TMLE-P, the initial category predictions
and their limits are positive almost surely.
The fitted fluctuation parameters converge to finite
population limits and solve their respective targeting
equations.
\end{enumerate}

In particular, these conditions require neither a finite-dimensional
outcome model nor derivatives of the outcome learner.

\subsubsection{Proof}
\label{app:proof-prop3}

We first show that replacing the fitted outcome predictions
by their limits has no first-order effect.
We then derive the contribution from propensity estimation
and transfer the resulting marginal expansion to the DOOR
trajectory and its weighted summaries.

\paragraph{Step 1: Replacing the outcome predictions by their limits}
Correct specification and the likelihood regularity
conditions imply
\[
P_0\bm s_\beta=\bm0,
\qquad
-P_0\left[
\frac{\partial\bm s_\beta(O;\bm\beta_0)}
{\partial\bm\beta^\top}\right]=\bm I_\beta.
\]
Expanding the training-sample likelihood equations gives
\begin{equation}
\widehat{\bm\beta}^{(-v)}-\bm\beta_0
=\bm I_\beta^{-1}P_{-v}\bm s_\beta
+o_p(n^{-1/2}),
\label{eq:propensity-mle-expansion}
\end{equation}
uniformly over the fixed number of folds.
The propensity derivative envelope and the
clipping condition give
\begin{equation}
\|\widehat\pi_a^{(-v)}-\pi_a\|_2
\leq
\|\pi_{\mathrm{raw}}(\cdot;\widehat{\bm\beta}^{(-v)})
-\pi\|_2
=O_p(n^{-1/2}).
\label{eq:propensity-rate}
\end{equation}

Recall \(\widetilde q_{at}^{\,r,v}\) from~\eqref{eq:qtilde}.
For TMLE and TMLE-P, positive limiting category
probabilities make the population targeting loss
strictly convex. Its finite solution is therefore
unique and common across folds.
Local Lipschitz continuity of the multinomial update,
together with consistency of the nuisance estimates
and fluctuation parameters, implies
\begin{equation}
\max_{a,t,k,v}
\|\widetilde q_{at}^{\,r,v}(k\mid\cdot)
-q_{at}^{r,\star}(k\mid\cdot)\|_2=o_p(1).
\label{eq:outcome-limit-convergence}
\end{equation}
For AIPW, this convergence follows directly from the
assumption on the initial outcome predictions.

Fix \(a,t,k\), and recall \(m_{at,k}\) from the
common marginal expansion in~\eqref{eq:marginal-decomposition}.
Replacing the fitted outcome prediction by its limit gives
\[
m_{at,k}(O,\widehat\pi_a^{(-v)},\widetilde q_{at}^{\,r,v})
-m_{at,k}(O,\widehat\pi_a^{(-v)},q_{at}^{r,\star})
=\left\{1-\frac{I(A=a)}{\widehat\pi_a^{(-v)}(X)}\right\}
\{\widetilde q_{at}^{\,r,v}(k\mid X)
-q_{at}^{r,\star}(k\mid X)\}.
\]
As in the proof of Theorem~\ref{thm:consistency-efficiency},
decompose its empirical average into a population term
and a centered empirical term.
By Cauchy--Schwarz and
\eqref{eq:propensity-rate}--\eqref{eq:outcome-limit-convergence},
the population term is
\(O_p(n^{-1/2})o_p(1)=o_p(n^{-1/2})\).
For AIPW, conditional cross-fitting gives the same
order for the centered term.
For TMLE and TMLE-P, first evaluate the fluctuation at
its fixed population limit, so that the same conditional
argument applies. The remaining difference is
\(o_p(n^{-1/2})\) by consistency of the fluctuation
parameters and the finite-dimensional empirical-process
argument used in the proof of
Theorem~\ref{thm:consistency-efficiency}.
For TMLE-P, the empirical supremum is taken before
substituting the shared fluctuation.
Consequently,
\[
\widehat p_{at,k}^{\,r}
=\sum_v\lambda_vP_vm_{at,k}
(O,\widehat\pi_a^{(-v)},\widetilde q_{at}^{\,r,v})
=\sum_v\lambda_vP_vm_{at,k}
(O,\widehat\pi_a^{(-v)},q_{at}^{r,\star})
+o_p(n^{-1/2}).
\]
Under the stated regularity conditions, this step
requires no convergence rate for the outcome predictions
or fluctuation parameters beyond consistency.

\paragraph{Step 2: Propensity-estimation correction}
With the outcome prediction fixed at \(q_{at}^{r,\star}\),
define
\[
\varphi_{at,k}^{r,\star}(O)
=m_{at,k}(O,\pi_a,q_{at}^{r,\star})-p_{at,k},
\qquad
\bm C_{at,k}^r
=P_0\{\varphi_{at,k}^{r,\star}(O)\bm s_\beta(O)^\top\}.
\]
Correct propensity specification and identification give
\(P_0\varphi_{at,k}^{r,\star}=0\).

To relate the propensity derivative to \(\bm C_{at,k}^r\),
write
\[
\dot{\bm\pi}(X)
=\left.\frac{\partial\pi_{\mathrm{raw}}(X;\bm\beta)}
{\partial\bm\beta}\right|_{\bm\beta=\bm\beta_0},
\qquad
\dot{\bm\pi}_a(X)=(2a-1)\dot{\bm\pi}(X).
\]
The Bernoulli likelihood score satisfies
\[
\bm s_\beta(O)
=\frac{A-\pi(X)}{\pi(X)\{1-\pi(X)\}}\dot{\bm\pi}(X).
\]
Since \(\E{\bm s_\beta(O)\mid X}=\bm0\), conditioning
on \(X\) gives
\[
\bm C_{at,k}^r
=P_0\left[
\frac{\dot{\bm\pi}_a(X)^\top}{\pi_a(X)}
\{q_{at}(k\mid X)-q_{at}^{r,\star}(k\mid X)\}
\right].
\]
Differentiating the augmented contribution and taking
expectations therefore yields
\[
P_0\left[
\left.
\frac{\partial m_{at,k}
(O,\pi_a(\cdot;\bm\beta),q_{at}^{r,\star})}
{\partial\bm\beta^\top}
\right|_{\bm\beta=\bm\beta_0}
\right]
=-P_0\left[
\frac{\dot{\bm\pi}_a(X)^\top}{\pi_a(X)}
\{q_{at}(k\mid X)-q_{at}^{r,\star}(k\mid X)\}
\right]
=-\bm C_{at,k}^r.
\]

A population Taylor expansion gives the linear term
\(-\bm C_{at,k}^r
(\widehat{\bm\beta}^{(-v)}-\bm\beta_0)\)
with remainder \(o_p(n^{-1/2})\).
Conditional cross-fitting and the root-\(n\) propensity
rate make the corresponding centered empirical difference
\(O_p(n^{-1})\).
Thus, for each fold,
\[
P_vm_{at,k}(O,\widehat\pi_a^{(-v)},q_{at}^{r,\star})
=P_vm_{at,k}(O,\pi_a,q_{at}^{r,\star})
-\bm C_{at,k}^r
(\widehat{\bm\beta}^{(-v)}-\bm\beta_0)
+o_p(n^{-1/2}).
\]
Combining this with Step 1 gives
\[
\begin{aligned}
\widehat p_{at,k}^{\,r}-p_{at,k}
&=P_n\varphi_{at,k}^{r,\star}
-\bm C_{at,k}^r\sum_v\lambda_v
(\widehat{\bm\beta}^{(-v)}-\bm\beta_0)
+o_p(n^{-1/2})\\
&=P_n\varphi_{at,k}^{r,\star}
-\bm C_{at,k}^r\bm I_\beta^{-1}
\frac1n\sum_{i=1}^n w_i\bm s_\beta(O_i)
+o_p(n^{-1/2}),
\end{aligned}
\]
where the second equality follows from
\eqref{eq:propensity-mle-expansion}, with
\(w_i=\sum_{v:i\notin I_v}\lambda_v/(1-\lambda_v)\)
accounting for each participant's inclusion in
multiple training samples.
Since \(\lambda_v\to1/V\), \(\max_i|w_i-1|\to0\).
The scores are centered, so a conditional variance
bound gives
\[
\frac1n\sum_{i=1}^n(w_i-1)\bm s_\beta(O_i)
=o_p(n^{-1/2}).
\]
Substitution yields the corrected marginal expansion
\[
\widehat p_{at,k}^{\,r}-p_{at,k}
=P_n\{\varphi_{at,k}^{r,\star}
-\bm C_{at,k}^r\bm I_\beta^{-1}\bm s_\beta\}
+o_p(n^{-1/2}).
\]

\paragraph{Step 3: DOOR expansion and variance}
Applying the bilinear expansion as in
Appendix~\ref{app:proof-prop2} gives
\[
\widehat{\bm\psi}^{\,r}-\bm\psi
=P_n\bm\phi^{r,\mathrm{corr}}+o_p(n^{-1/2}),
\qquad
\widehat\Psi_\omega^{\,r}-\Psi_\omega
=P_n\phi_\omega^{r,\mathrm{corr}}+o_p(n^{-1/2}),
\]
where
\[
\bm\phi^{r,\mathrm{corr}}
=\bm\phi^{r,\star}-\bm C^r\bm I_\beta^{-1}\bm s_\beta,
\qquad
\bm C^r=P_0\{\bm\phi^{r,\star}\bm s_\beta^\top\},
\qquad
\phi_\omega^{r,\mathrm{corr}}
=\bm\omega^\top\bm\phi^{r,\mathrm{corr}}.
\]
The corrected contributions are centered and have finite
second moments. The central limit theorem and Slutsky's
theorem therefore yield
\[
\sqrt n(\widehat{\bm\psi}^{\,r}-\bm\psi)
\xrightarrow{d}N_T(\bm0,\bm\Sigma_r),
\qquad
\sqrt n(\widehat\Psi_\omega^{\,r}-\Psi_\omega)
\xrightarrow{d}N(0,\bm\omega^\top\bm\Sigma_r\bm\omega),
\]
where, by the definition of \(\bm C^r\) and
\(P_0(\bm s_\beta\bm s_\beta^\top)=\bm I_\beta\),
\[
\bm\Sigma_r
=\operatorname{Var}(\bm\phi^{r,\star})
-\bm C^r\bm I_\beta^{-1}(\bm C^r)^\top
\preceq\operatorname{Var}(\bm\phi^{r,\star}).
\]
Multiplying by \(\bm\omega^\top\) and \(\bm\omega\)
gives the corresponding weighted-summary variance identity.

For the fitted corrections described in the implementation
details, consistency of the estimated matrices, the score
envelope, the outcome and targeting convergence above,
and \(\max_i|w_i-1|\to0\) give
\[
\frac1n\sum_{i=1}^n
\|\widehat{\bm\phi}_i^{\,r,\mathrm{corr}}
-\bm\phi^{r,\mathrm{corr}}(O_i)\|^2=o_p(1).
\]
The covariance-consistency argument in
Appendix~\ref{app:proof-prop2} then applies, giving
\(n\widehat V_r\xrightarrow{p}\bm\Sigma_r\) and
\(n\bm\omega^\top\widehat V_r\bm\omega
\xrightarrow{p}\bm\omega^\top\bm\Sigma_r\bm\omega\).
\hfill\(\square\)

\appendixsection{Implementation Details}
\label{app:simulation}

\subsection{G-computation and robust U-statistic estimators}
\label{app:gu}

For G-computation, the cross-fitted marginal category probabilities and win probability are
\[
\widehat p_{at,k}^{\,G}
=
\sum_{v=1}^V \frac{n_v}{n}
P_v \widehat q_{at}^{(-v)}(k\mid X),
\qquad
\widehat\psi_t^{\,G}
=
(\widehat{\bm p}_{1t}^{\,G})^\top
W\widehat{\bm p}_{0t}^{\,G}.
\]
The non-cross-fitted estimator replaces the fold-specific predictions by full-sample predictions and averages them using \(P_n\). In both cases, the category marginals are averaged before applying the win-probability contrast.

For the U-statistic comparator, define
\[
\mu_t(x,x')
=
\sum_{k=1}^K \sum_{\ell=1}^K
w_{k\ell}
q_{1t}(k\mid x)
q_{0t}(\ell\mid x').
\]
For distinct participants \(i,j\in I_v\), let
\[
\widehat m_{t,ij}^{\,v}
=
\widehat\mu_t^{(-v)}(X_i,X_j)+
\frac{A_i(1-A_j)}
{\widehat\pi^{(-v)}(X_i)
 \{1-\widehat\pi^{(-v)}(X_j)\}}
\left\{
w_{Y_i(t),Y_j(t)}
-
\widehat\mu_t^{(-v)}(X_i,X_j)
\right\},
\]
where \(\widehat\mu_t^{(-v)}\) is constructed from the two outcome regressions \(\widehat q_{at}^{(-v)}\) trained outside \(I_v\). The within-fold estimator and its pooled version are
\[
\widehat\psi_t^{\,U,v}
=
\binom{n_v}{2}^{-1}
\sum_{i<j,\ i,j\in I_v}
\frac{\widehat m_{t,ij}^{\,v}
      +\widehat m_{t,ji}^{\,v}}{2},
\qquad
\widehat\psi_t^{\,U}
=
\sum_{v=1}^V
\frac{n_v}{n}\widehat\psi_t^{\,U,v}.
\]
Cross-fitting is less direct for the pairwise estimator because both participants in each evaluated pair must be excluded from nuisance training. Our implementation uses only within-fold pairs, retaining approximately \(1/V\) of all distinct pairs when the \(V\) folds are equally sized. Including between-fold pairs could instead be accommodated by training nuisance models outside each pair of folds, but this would require additional model fits. The non-cross-fitted version uses all distinct pairs and full-sample nuisance fits. 

\subsection{Data-generating mechanism}

We generate five mutually independent baseline covariates:
\[
X_1\sim\operatorname{Bernoulli}(0.2),
\qquad
X_2\sim\operatorname{Bernoulli}(0.6), 
\quad
\Prb{X_3=A,B,C}=(0.40,0.35,0.25),
\]
\[
X_4\sim N(-1,1),
\qquad
X_5\sim N(1,1).
\]
Treatment is then generated as
\(A\mid\boldsymbol{X}\sim\operatorname{Bernoulli}\{\pi(\boldsymbol{X})\}\),
where
\[
\operatorname{logit}\{\pi(\boldsymbol{X})\}
=
s\bigl\{
0.8+0.8X_1-0.5X_2
+0.4I(X_3=\mathrm{B})-0.4I(X_3=\mathrm{C})
+X_4+0.3X_5
\bigr\}.
\]
The scale parameter \(s\) controls treatment overlap while preserving
the logistic model form. We use \(s=1\) in Regimes 1--4 and
\(s\in\{1.5,2\}\) in Regime 5, with larger values producing weaker
overlap. Changing \(s\) leaves the covariate and potential-outcome
distributions unchanged and therefore does not alter the target win
probabilities.

We generate potential outcomes with four ordered categories using an
arm-specific latent probit model. For treatment arm \(a\in\{0,1\}\)
and visit \(j=1,\ldots,6\), define
\[
Y_j(a)=k
\quad\text{if}\quad
\alpha_{k-1}<Z_j(a)\leq\alpha_k,
\qquad k=1,\ldots,4,
\]
with cutpoints
\[
(\alpha_0,\ldots,\alpha_4)
=(-\infty,-1,0,1,\infty)
\]
and latent variable
\[
Z_j(a)
=-\eta_a(\boldsymbol{X})-\delta_{aj}+\varepsilon_{aj},
\]
where \(\eta_a(\boldsymbol{X})\) captures baseline covariate effects,
\(\delta_{aj}\) specifies the visit-specific shift, and the correlated
errors \(\varepsilon_{aj}\) induce within-participant dependence
across visits. The arm-specific covariate predictors are chosen to be
\begin{align*}
\eta_0(\boldsymbol{X})
&=0.2+0.2X_1+0.1X_2
  +0.3I(X_3=\mathrm{B})-0.3I(X_3=\mathrm{C})
  +1.2X_4+0.7X_5,\\
\eta_1(\boldsymbol{X})
&=0.5+0.5X_1+0.3X_2
  +0.25I(X_3=\mathrm{B})-0.25I(X_3=\mathrm{C})
  +0.9X_4-0.5X_5.
\end{align*}
The shifts across the six visits are
\[
\boldsymbol{\delta}_0=(0,0.12,0.24,0.36,0.48,0.60),
\qquad
\boldsymbol{\delta}_1=(0,0.08,0.16,0.24,0.32,0.40).
\]
For each treatment arm, we generate
\[
\boldsymbol{\varepsilon}_a
=(\varepsilon_{a1},\ldots,\varepsilon_{a6})^\top
\sim N_6(\boldsymbol{0},\boldsymbol{\Sigma}),
\qquad
\Sigma_{jk}=0.6^{|j-k|}.
\]
The error vectors are generated independently of \(\boldsymbol X\)
and of each other, and treatment is generated independently of these
errors conditional on \(\boldsymbol X\).
The observed outcome is \(Y(t_j)=Y_j(A)\).
Consequently, the conditional category probabilities are
\[
q_{a,t_j}(k\mid\boldsymbol{X})
=
\Phi\{\alpha_k+\eta_a(\boldsymbol{X})+\delta_{aj}\}
-
\Phi\{\alpha_{k-1}+\eta_a(\boldsymbol{X})+\delta_{aj}\}.
\]

We approximate \(p_{a,t_j,k}\) by averaging these known conditional
probabilities over \(M=100{,}000\) independently generated covariate
vectors, and calculate $\psi_{t_j} =\boldsymbol{p}_{1,t_j}^{\top}W\boldsymbol{p}_{0,t_j}$.
The resulting reference value for the equally weighted average,
\(\Psi_\omega=6^{-1}\sum_{j=1}^6\psi_{t_j}\), is 0.585791.
The same reference values are used for both sample sizes and all regimes.

\subsection{Instantiation of the inference corrections}
\label{app:instantiation}

We give the matrices and fitted contributions for full-sample
AIPW and TMLE, cross-fitted AIPW and TMLE, and TMLE-P,
followed by the simpler correction under a correctly specified
propensity model. The cross-fitted constructions follow
Propositions~\ref{prop:corrected-parametric}
and~\ref{prop:correct-ps}. The full-sample versions use
\(P_n\) throughout and require the corresponding
full-sample regularity conditions.
When the working propensity is fixed, its parameter, score, and derivative blocks are omitted.

\paragraph{Common matrix blocks}
For fold \(v\), let \(\bm\eta_v\) collect the propensity
parameters and the outcome-model parameters for all
treatment arms and visits, with unnormalized score
\(\bm s_\eta(O;\bm\eta_v)\).
Fix an arm \(a\) and visit \(t\), and write its unnormalized targeting score as
\(\bm s_{\epsilon,at}
(O;\bm\eta_v,\bm\epsilon_{at}^{r,v})\),
where \(\bm\epsilon_{at}^{r,v}\) is fold-specific for
TMLE and shared across folds for TMLE-P.
Define
\[
\bm A_{\eta,v}
=P_0\left[
\frac{\partial\bm s_\eta(O;\bm\eta_v)}
{\partial\bm\eta_v^\top}
\right],
\]
\[
\bm A_{at,\epsilon\eta,v}
=P_0\left[
\frac{\partial\bm s_{\epsilon,at}
(O;\bm\eta_v,\bm\epsilon_{at}^{r,v})}
{\partial\bm\eta_v^\top}
\right],
\qquad
\bm A_{at,\epsilon,v}
=P_0\left[
\frac{\partial\bm s_{\epsilon,at}
(O;\bm\eta_v,\bm\epsilon_{at}^{r,v})}
{\partial(\bm\epsilon_{at}^{r,v})^\top}
\right].
\]
For AIPW, write
\[
\bm m_{at}(O;\bm\eta_v)
=\bm q_{at}(X;\bm\eta_v)
+\frac{I(A=a)}{\pi_a(X;\bm\beta_v)}
\{\bm z_t-\bm q_{at}(X;\bm\eta_v)\},
\qquad
\bm G_{at,\eta,v}^{\mathrm{AIPW}}
=P_0\left[
\frac{\partial\bm m_{at}(O;\bm\eta_v)}
{\partial\bm\eta_v^\top}
\right].
\]
For the targeted estimators, define
\[
\bm G_{at,\eta,v}
=P_0\left[
\frac{\partial\bm q_{at}^{\bm\epsilon_{at}^{r,v}}
(X;\bm\eta_v)}
{\partial\bm\eta_v^\top}
\right],
\qquad
\bm G_{at,\epsilon,v}
=P_0\left[
\frac{\partial\bm q_{at}^{\bm\epsilon_{at}^{r,v}}
(X;\bm\eta_v)}
{\partial(\bm\epsilon_{at}^{r,v})^\top}
\right].
\]
For a fixed \((a,t)\), only the columns corresponding
to \(\bm\beta_v\) and \(\bm\theta_{at,v}\) can be nonzero
in \(\bm G_{at,\eta,v}^{\mathrm{AIPW}}\),
\(\bm G_{at,\eta,v}\), and
\(\bm A_{at,\epsilon\eta,v}\).
Below, replacing the subscript \(\eta\) by \(\beta\)
or \(\theta_{at}\) denotes the corresponding column
block, and the same convention applies to other matrices.
All population blocks are evaluated at the corresponding
parameter limits. Their fitted versions use the training
average \(P_{-v}\) for \(\widehat{\bm A}_{\eta,v}\) and
the validation average \(P_v\) for every other block.
Without cross-fitting, all blocks use \(P_n\) and the
index \(v\) is omitted. In the stacked matrices below,
hats indicate these empirical estimates. Under the stated regularity
conditions, these empirical matrices converge to
\(\bm A\) and \(\bm G_{at}\).

The initial propensity and arm--visit outcome models
are fitted separately, giving
\[
\widehat{\bm A}_{\eta,v}
=\operatorname{blockdiag}\left(
\widehat{\bm A}_{\beta,v},
\{\widehat{\bm A}_{\theta_{at},v}\}
_{a\in\{0,1\},\,t=1,\ldots,T}
\right).
\]
Each block is estimated by averaging the corresponding
score derivatives at the fitted parameters, as given in
Appendix~\ref{app:implementation-derivatives}.
The training averages use all \(n-n_v\) training participants,
with arm indicators included in the outcome scores.
Using these blocks, we give the matrices and corrected
contributions for each of the five estimators below.

\paragraph{1. Full-sample AIPW}
Here \(\bm\gamma=\bm\eta\), so
\[
\widehat{\bm A}=\widehat{\bm A}_\eta,
\qquad
\widehat{\bm G}_{at}
=\widehat{\bm G}_{at,\eta}^{\mathrm{AIPW}},
\qquad
\widehat{\bm\varphi}_{at,i}^{\mathrm{AIPW},\mathrm{corr}}
=\widehat{\bm m}_{at,i}
-\widehat{\bm p}_{at}^{\mathrm{AIPW}}
-\widehat{\bm G}_{at,\eta}^{\mathrm{AIPW}}
\widehat{\bm A}_\eta^{-1}
\bm s_\eta(O_i;\widehat{\bm\eta}),
\]
where
\(\widehat{\bm m}_{at,i}
=\bm m_{at}(O_i;\widehat{\bm\eta})\)
uses the full-sample nuisance fit.
Equivalently, the blockwise expression can be derived as
\[
\begin{aligned}
\widehat{\bm\varphi}_{at,i}^{\mathrm{AIPW},\mathrm{corr}}
&=\widehat{\bm m}_{at,i}-\widehat{\bm p}_{at}^{\mathrm{AIPW}}-\widehat{\bm G}_{at,\beta}^{\mathrm{AIPW}}
\widehat{\bm A}_{\beta}^{-1}
\bm s_{\beta}(O_i;\widehat{\bm\beta})
-\widehat{\bm G}_{at,\theta_{at}}^{\mathrm{AIPW}}
\widehat{\bm A}_{\theta_{at}}^{-1}
\bm s_{\theta_{at}}(O_i;\widehat{\bm\theta}_{at}).
\end{aligned}
\]
\paragraph{2. Full-sample TMLE}
For the marginal at \((a,t)\), the relevant parameters
are the initial nuisance parameters \(\bm\eta\) and
the fluctuation parameter \(\bm\epsilon_{at}\).
The corresponding matrices are
\[
\widehat{\bm A}
=\begin{pmatrix}
\widehat{\bm A}_\eta & \bm0\\
\widehat{\bm A}_{at,\epsilon\eta}
& \widehat{\bm A}_{at,\epsilon}
\end{pmatrix},
\qquad
\widehat{\bm G}_{at}
=(\widehat{\bm G}_{at,\eta},
\widehat{\bm G}_{at,\epsilon}).
\]
Define the targeting matrices
\[
\widehat{\bm B}_{at}
=\widehat{\bm G}_{at,\epsilon}
\widehat{\bm A}_{at,\epsilon}^{-1},
\qquad
\widehat{\bm L}_{at,\eta}
=\widehat{\bm G}_{at,\eta}
-\widehat{\bm B}_{at}
\widehat{\bm A}_{at,\epsilon\eta}.
\]
Block inversion then gives
\[
\widehat{\bm\varphi}_{at,i}^{\mathrm{TMLE},\mathrm{corr}}
=\bm q_{at}^{\widehat{\bm\epsilon}_{at}}
(X_i;\widehat{\bm\eta})
-\widehat{\bm p}_{at}^{\mathrm{TMLE}}
-\widehat{\bm B}_{at}
\bm s_{\epsilon,at}
(O_i;\widehat{\bm\eta},\widehat{\bm\epsilon}_{at})
-\widehat{\bm L}_{at,\eta}
\widehat{\bm A}_\eta^{-1}
\bm s_\eta(O_i;\widehat{\bm\eta}).
\]
Equivalently, the blockwise expression is
\[
\begin{aligned}
\widehat{\bm\varphi}_{at,i}^{\mathrm{TMLE},\mathrm{corr}}
&=\bm q_{at}^{\widehat{\bm\epsilon}_{at}}(X_i;\widehat{\bm\eta})-\widehat{\bm p}_{at}^{\mathrm{TMLE}}\\
&\quad -\widehat{\bm B}_{at}\bm s_{\epsilon,at}(O_i;\widehat{\bm\eta},\widehat{\bm\epsilon}_{at}) -\widehat{\bm L}_{at,\beta}
\widehat{\bm A}_{\beta}^{-1}
\bm s_{\beta}(O_i;\widehat{\bm\beta})\\
&\quad -\widehat{\bm L}_{at,\theta_{at}}
\widehat{\bm A}_{\theta_{at}}^{-1}
\bm s_{\theta_{at}}(O_i;\widehat{\bm\theta}_{at}).
\end{aligned}
\]

\paragraph{3. Cross-fitted AIPW}
Stacking the initial nuisance parameters across folds gives
\[
\widehat{\bm A}
=\operatorname{blockdiag}_{v=1}^V
(\widehat{\bm A}_{\eta,v}),
\qquad
\widehat{\bm G}_{at}
=(\lambda_1\widehat{\bm G}_{at,\eta,1}^{\mathrm{AIPW}},
\ldots,
\lambda_V\widehat{\bm G}_{at,\eta,V}^{\mathrm{AIPW}}).
\]
The training-score block in \(\bm s_i(\bm\gamma)\),
defined in Appendix~\ref{app:inference-definitions},
is \(I(i\notin I_v)\bm s_\eta(O_i;\bm\eta_v)/(1-\lambda_v)\).
Combining this normalization with the factor \(\lambda_v\)
in the marginal derivative gives
\[
\widehat{\bm\varphi}_{at,i}^{\mathrm{AIPW},\mathrm{corr}}
=\widehat{\bm m}_{at,i}
-\widehat{\bm p}_{at}^{\mathrm{AIPW}}
-\sum_{v:i\notin I_v}\frac{\lambda_v}{1-\lambda_v}
\widehat{\bm G}_{at,\eta,v}^{\mathrm{AIPW}}
\widehat{\bm A}_{\eta,v}^{-1}
\bm s_\eta(O_i;\widehat{\bm\eta}_v).
\]
where
\(\widehat{\bm m}_{at,i}
=\bm m_{at}(O_i;\widehat{\bm\eta}_{v(i)})\),
and each score correction uses a training sample
containing participant \(i\).
Equivalently, the implemented blockwise expression is
\[
\begin{aligned}
\widehat{\bm\varphi}_{at,i}^{\mathrm{AIPW},\mathrm{corr}}
&=\widehat{\bm m}_{at,i}-\widehat{\bm p}_{at}^{\mathrm{AIPW}}\\
&\quad -\sum_{v:i\notin I_v}\frac{\lambda_v}{1-\lambda_v}\Bigl\{
\widehat{\bm G}_{at,\beta,v}^{\mathrm{AIPW}}
\widehat{\bm A}_{\beta,v}^{-1}
\bm s_{\beta}(O_i;\widehat{\bm\beta}_{v})\\
&\qquad\qquad\qquad +\widehat{\bm G}_{at,\theta_{at},v}^{\mathrm{AIPW}}
\widehat{\bm A}_{\theta_{at},v}^{-1}
\bm s_{\theta_{at}}(O_i;\widehat{\bm\theta}_{at,v})\Bigr\}.
\end{aligned}
\]

\paragraph{4. Cross-fitted TMLE}
Ordering the parameters as
\((\bm\eta_1,\bm\epsilon_{at}^1,\ldots,
\bm\eta_V,\bm\epsilon_{at}^V)\) gives
\[
\widehat{\bm A}
=\operatorname{blockdiag}_{v=1}^V
\begin{pmatrix}
\widehat{\bm A}_{\eta,v} & \bm0\\
\widehat{\bm A}_{at,\epsilon\eta,v}
& \widehat{\bm A}_{at,\epsilon,v}
\end{pmatrix},
\qquad
\widehat{\bm G}_{at}
=\bigl(\lambda_v\widehat{\bm G}_{at,\eta,v},
\lambda_v\widehat{\bm G}_{at,\epsilon,v}\bigr)_{v=1}^V.
\]
Set
\(\widehat{\bm B}_{at,v}
=\widehat{\bm G}_{at,\epsilon,v}
\widehat{\bm A}_{at,\epsilon,v}^{-1}\)
and
\(\widehat{\bm L}_{at,\eta,v}
=\widehat{\bm G}_{at,\eta,v}
-\widehat{\bm B}_{at,v}
\widehat{\bm A}_{at,\epsilon\eta,v}\).
Substituting these blocks into the corrected marginal
contribution gives
\[
\begin{aligned}
\widehat{\bm\varphi}_{at,i}^{\mathrm{TMLE},\mathrm{corr}}
&=\bm q_{at}^{\widehat{\bm\epsilon}_{at}^{\mathrm{TMLE},v(i)}}
(X_i;\widehat{\bm\eta}_{v(i)})
-\widehat{\bm p}_{at}^{\mathrm{TMLE}}
-\widehat{\bm B}_{at,v(i)}
\bm s_{\epsilon,at}
(O_i;\widehat{\bm\eta}_{v(i)},
\widehat{\bm\epsilon}_{at}^{\mathrm{TMLE},v(i)})\\
&\quad-\sum_{v:i\notin I_v}\frac{\lambda_v}{1-\lambda_v}
\widehat{\bm L}_{at,\eta,v}
\widehat{\bm A}_{\eta,v}^{-1}
\bm s_\eta(O_i;\widehat{\bm\eta}_v).
\end{aligned}
\]
Equivalently, the implemented blockwise expression is
\[
\begin{aligned}
\widehat{\bm\varphi}_{at,i}^{\mathrm{TMLE},\mathrm{corr}}
&=\bm q_{at}^{\widehat{\bm\epsilon}_{at}^{\mathrm{TMLE},v(i)}}(X_i;\widehat{\bm\eta}_{v(i)})-\widehat{\bm p}_{at}^{\mathrm{TMLE}} -\widehat{\bm B}_{at,v(i)}\bm s_{\epsilon,at}(O_i;\widehat{\bm\eta}_{v(i)},\widehat{\bm\epsilon}_{at}^{\mathrm{TMLE},v(i)})\\
&\quad-\sum_{v:i\notin I_v}\frac{\lambda_v}{1-\lambda_v}\Bigl\{
\widehat{\bm L}_{at,\beta,v}
\widehat{\bm A}_{\beta,v}^{-1}
\bm s_{\beta}(O_i;\widehat{\bm\beta}_{v})
+\widehat{\bm L}_{at,\theta_{at},v}
\widehat{\bm A}_{\theta_{at},v}^{-1}
\bm s_{\theta_{at}}(O_i;\widehat{\bm\theta}_{at,v})\Bigr\}.
\end{aligned}
\]

\paragraph{5. TMLE-P}
The initial nuisance parameters remain fold-specific,
but \(\bm\epsilon_{at}\) is shared. With ordering
\((\bm\eta_1,\ldots,\bm\eta_V,\bm\epsilon_{at})\),
the matrices are
\[
\widehat{\bm A}
=\begin{pmatrix}
\widehat{\bm A}_{\eta,1} & \cdots & \bm0 & \bm0\\
\vdots & \ddots & \vdots & \vdots\\
\bm0 & \cdots & \widehat{\bm A}_{\eta,V} & \bm0\\
\lambda_1\widehat{\bm A}_{at,\epsilon\eta,1}
& \cdots &
\lambda_V\widehat{\bm A}_{at,\epsilon\eta,V}
& \sum_v\lambda_v\widehat{\bm A}_{at,\epsilon,v}
\end{pmatrix},
\]
\[
\widehat{\bm G}_{at}
=\left(
\lambda_1\widehat{\bm G}_{at,\eta,1},\ldots,
\lambda_V\widehat{\bm G}_{at,\eta,V},
\sum_v\lambda_v\widehat{\bm G}_{at,\epsilon,v}
\right).
\]
All targeting blocks are evaluated at the shared fitted
fluctuation. Define
\[
\widehat{\bm B}_{at,P}
=\left(\sum_v\lambda_v
\widehat{\bm G}_{at,\epsilon,v}\right)
\left(\sum_v\lambda_v
\widehat{\bm A}_{at,\epsilon,v}\right)^{-1},
\qquad
\widehat{\bm L}_{at,\eta,v}^{P}
=\widehat{\bm G}_{at,\eta,v}
-\widehat{\bm B}_{at,P}
\widehat{\bm A}_{at,\epsilon\eta,v}.
\]
Substituting these blocks into the corrected marginal
contribution gives
\[
\begin{aligned}
\widehat{\bm\varphi}_{at,i}^{\mathrm{TMLE\text{-}P},\mathrm{corr}}
&=\bm q_{at}^{\widehat{\bm\epsilon}_{at}^{\mathrm{TMLE\text{-}P}}}
(X_i;\widehat{\bm\eta}_{v(i)})
-\widehat{\bm p}_{at}^{\mathrm{TMLE\text{-}P}}
-\widehat{\bm B}_{at,P}
\bm s_{\epsilon,at}
(O_i;\widehat{\bm\eta}_{v(i)},
\widehat{\bm\epsilon}_{at}^{\mathrm{TMLE\text{-}P}})\\
&\quad-\sum_{v:i\notin I_v}\frac{\lambda_v}{1-\lambda_v}
\widehat{\bm L}_{at,\eta,v}^{P}
\widehat{\bm A}_{\eta,v}^{-1}
\bm s_\eta(O_i;\widehat{\bm\eta}_v).
\end{aligned}
\]
Equivalently, the implemented blockwise expression is
\[
\begin{aligned}
\widehat{\bm\varphi}_{at,i}^{\mathrm{TMLE\text{-}P},\mathrm{corr}}
&=\bm q_{at}^{\widehat{\bm\epsilon}_{at}^{\mathrm{TMLE\text{-}P}}}(X_i;\widehat{\bm\eta}_{v(i)})-\widehat{\bm p}_{at}^{\mathrm{TMLE\text{-}P}}-\widehat{\bm B}_{at,P}\bm s_{\epsilon,at}(O_i;\widehat{\bm\eta}_{v(i)},\widehat{\bm\epsilon}_{at}^{\mathrm{TMLE\text{-}P}})\\
&\quad-\sum_{v:i\notin I_v}\frac{\lambda_v}{1-\lambda_v}\Bigl\{
\widehat{\bm L}_{at,\beta,v}^{P}
\widehat{\bm A}_{\beta,v}^{-1}
\bm s_{\beta}(O_i;\widehat{\bm\beta}_{v})
+\widehat{\bm L}_{at,\theta_{at},v}^{P}
\widehat{\bm A}_{\theta_{at},v}^{-1}
\bm s_{\theta_{at}}(O_i;\widehat{\bm\theta}_{at,v})\Bigr\}.
\end{aligned}
\]

The targeting Jacobians and marginal derivatives
are pooled before inversion.

\paragraph{Propensity information and the simpler correction}
Under Proposition~\ref{prop:correct-ps}, the code estimates
the propensity information by the negative training-sample
likelihood Hessian,
\[
\widehat{\bm I}_{\beta,v}
=-\widehat{\bm A}_{\beta,v}.
\]
For the logistic fit, this is the negative Hessian based
on raw model probabilities, given in
Appendix~\ref{app:implementation-derivatives}.
Let \(\widehat{\bm\phi}_i^{r,\star}\) be the augmented
trajectory contribution, using the initial predictions
for AIPW and targeted predictions for TMLE or TMLE-P.
The covariance matrix $C_v^r$ is estimated from
\[
\widehat{\bm C}_v^r
=P_v\left[
(\widehat{\bm\phi}^{r,\star}
-P_v\widehat{\bm\phi}^{r,\star})
\{\bm s_\beta(O;\widehat{\bm\beta}_v)
-P_v\bm s_\beta(O;\widehat{\bm\beta}_v)\}^\top
\right]
\]
and forms
\[
\widehat{\bm\phi}_i^{r,\mathrm{corr}}
=\widehat{\bm\phi}_i^{r,\star}
-\sum_{v:i\notin I_v}\frac{\lambda_v}{1-\lambda_v}
\widehat{\bm C}_v^r\widehat{\bm I}_{\beta,v}^{-1}
\{\bm s_\beta(O_i;\widehat{\bm\beta}_v)
-P_{-v}\bm s_\beta(O;\widehat{\bm\beta}_v)\}.
\]
Thus each validation-fold covariance is paired with
its corresponding training-sample information matrix.
For a full-sample fit, both matrices use \(P_n\), and
\[
\widehat{\bm\phi}_i^{r,\mathrm{corr}}
=\widehat{\bm\phi}_i^{r,\star}
-\widehat{\bm C}^r\widehat{\bm I}_\beta^{-1}
\{\bm s_\beta(O_i;\widehat{\bm\beta})
-P_n\bm s_\beta(O;\widehat{\bm\beta})\}.
\]

\paragraph{DOOR contributions and implementation conventions}
For the complete correction, the code combines the
corrected marginal vectors using
\[
\widehat\phi_{t,i}^{r,\mathrm{corr}}
=(\widehat{\bm\varphi}_{1t,i}^{r,\mathrm{corr}})^\top
W\widehat{\bm p}_{0t}^r
+(\widehat{\bm p}_{1t}^r)^\top
W\widehat{\bm\varphi}_{0t,i}^{r,\mathrm{corr}}.
\]
All cross-fitted methods use pooled marginal estimates
in these coefficients.
The implemented initial likelihood scores and Hessians
use raw model probabilities. Estimation and targeting
use clipped propensity predictions and clipped,
renormalized category predictions.
The complete correction is implemented for ordinal
probit and multinomial logistic outcome models, with
derivatives given in
Appendix~\ref{app:implementation-derivatives}.
The simpler correct-propensity correction uses no
outcome derivatives.

\subsection{Derivatives}
\label{app:implementation-derivatives}

We give the derivatives needed to construct the blocks in
Appendix~\ref{app:instantiation}. All derivatives are
evaluated at the fitted parameters when computing the
empirical matrices.

\paragraph{Logistic propensity model}
For
\(\pi_{\mathrm{raw}}(X;\bm\beta)
=\operatorname{expit}\{\bm x_\pi(X)^\top\bm\beta\}\),
\[
\frac{\partial\pi_{\mathrm{raw}}(X;\bm\beta)}
{\partial\bm\beta}
=\pi_{\mathrm{raw}}(X;\bm\beta)
\{1-\pi_{\mathrm{raw}}(X;\bm\beta)\}\bm x_\pi(X).
\]
The likelihood score and its Jacobian are
\[
\begin{aligned}
\bm s_\beta(O;\bm\beta)
&=\bm x_\pi(X)\{A-\pi_{\mathrm{raw}}(X;\bm\beta)\},\\
\frac{\partial\bm s_\beta(O;\bm\beta)}
{\partial\bm\beta^\top}
&=-\pi_{\mathrm{raw}}(X;\bm\beta)
\{1-\pi_{\mathrm{raw}}(X;\bm\beta)\}
\bm x_\pi(X)\bm x_\pi(X)^\top.
\end{aligned}
\]
Thus,
\[
\widehat{\bm A}_{\beta,v}
=P_{-v}\left[
\left.
\frac{\partial\bm s_\beta(O;\bm\beta)}
{\partial\bm\beta^\top}
\right|_{\bm\beta=\widehat{\bm\beta}_v}
\right],
\qquad
\widehat{\bm I}_{\beta,v}
=-\widehat{\bm A}_{\beta,v}.
\]

\paragraph{Augmented contributions and multinomial targeting}
For the augmented contribution \(\bm m_{at}(O;\bm\eta)\)
defined in Appendix~\ref{app:instantiation}, the derivatives are
\[
\frac{\partial\bm m_{at}(O;\bm\eta)}
{\partial\bm\theta_{at}^\top}
=\left\{1-\frac{I(A=a)}{\pi_a(X;\bm\beta)}\right\}
\frac{\partial\bm q_{at}(X;\bm\theta_{at})}
{\partial\bm\theta_{at}^\top},
\]
\[
\frac{\partial\bm m_{at}(O;\bm\eta)}
{\partial\bm\beta^\top}
=-\frac{I(A=a)}{\pi_a(X;\bm\beta)^2}
\{\bm z_t-\bm q_{at}(X;\bm\theta_{at})\}
\frac{\partial\pi_a(X;\bm\beta)}
{\partial\bm\beta^\top}.
\]
For the logistic model, the implementation uses
\[
\frac{\partial\pi_a(X;\bm\beta)}
{\partial\bm\beta}
=(2a-1)\pi(X;\bm\beta)
\{1-\pi(X;\bm\beta)\}\bm x_\pi(X)
\]
as its smooth prediction derivative.

For the multinomial fluctuation, let
\(\bm E=(\bm e_1,\ldots,\bm e_{K-1})\) and define
\[
\bm Q_{at}^{\bm\epsilon}(X;\bm\eta)
=\operatorname{diag}\{
\bm q_{at}^{\bm\epsilon}(X;\bm\eta)\}
-\bm q_{at}^{\bm\epsilon}(X;\bm\eta)
\bm q_{at}^{\bm\epsilon}(X;\bm\eta)^\top.
\]
The prediction derivatives are
\[
\frac{\partial\bm q_{at}^{\bm\epsilon}(X;\bm\eta)}
{\partial\bm\theta_{at}^\top}
=\bm Q_{at}^{\bm\epsilon}(X;\bm\eta)
\operatorname{diag}\{\bm q_{at}(X;\bm\theta_{at})\}^{-1}
\frac{\partial\bm q_{at}(X;\bm\theta_{at})}
{\partial\bm\theta_{at}^\top},
\]
\[
\frac{\partial\bm q_{at}^{\bm\epsilon}(X;\bm\eta)}
{\partial\bm\beta^\top}
=-\frac{\bm Q_{at}^{\bm\epsilon}(X;\bm\eta)
\bm E\bm\epsilon}{\pi_a(X;\bm\beta)^2}
\frac{\partial\pi_a(X;\bm\beta)}
{\partial\bm\beta^\top},
\qquad
\frac{\partial\bm q_{at}^{\bm\epsilon}(X;\bm\eta)}
{\partial\bm\epsilon^\top}
=\frac{\bm Q_{at}^{\bm\epsilon}(X;\bm\eta)\bm E}
{\pi_a(X;\bm\beta)}.
\]
The targeting score is
\[
\bm s_{\epsilon,at}(O;\bm\eta,\bm\epsilon)
=\frac{I(A=a)}{\pi_a(X;\bm\beta)}
\bm E^\top\{\bm z_t-\bm q_{at}^{\bm\epsilon}(X;\bm\eta)\}.
\]
Its derivatives are
\[
\begin{aligned}
\frac{\partial\bm s_{\epsilon,at}(O;\bm\eta,\bm\epsilon)}
{\partial\bm\beta^\top}
&=-\frac{I(A=a)}{\pi_a(X;\bm\beta)^2}
\bm E^\top\{\bm z_t-\bm q_{at}^{\bm\epsilon}(X;\bm\eta)\}
\frac{\partial\pi_a(X;\bm\beta)}{\partial\bm\beta^\top}\\
&\quad-\frac{I(A=a)}{\pi_a(X;\bm\beta)}
\bm E^\top
\frac{\partial\bm q_{at}^{\bm\epsilon}(X;\bm\eta)}
{\partial\bm\beta^\top},
\end{aligned}
\]
\[
\begin{aligned}
\frac{\partial\bm s_{\epsilon,at}(O;\bm\eta,\bm\epsilon)}
{\partial\bm\theta_{at}^\top}
&=-\frac{I(A=a)}{\pi_a(X;\bm\beta)}
\bm E^\top
\frac{\partial\bm q_{at}^{\bm\epsilon}(X;\bm\eta)}
{\partial\bm\theta_{at}^\top},\\
\frac{\partial\bm s_{\epsilon,at}(O;\bm\eta,\bm\epsilon)}
{\partial\bm\epsilon^\top}
&=-\frac{I(A=a)}{\pi_a(X;\bm\beta)^2}
\bm E^\top\bm Q_{at}^{\bm\epsilon}(X;\bm\eta)\bm E.
\end{aligned}
\]
Averaging the score and prediction derivatives over
the validation sample gives the targeting Jacobian
and marginal derivative blocks, respectively.
Without cross-fitting, the full sample is directly used.

\paragraph{Ordinal probit outcome model}
Let \(\bm b_{at}\) denote the slopes and
\(\zeta_{at,1}<\cdots<\zeta_{at,K-1}\) the cutpoints,
with \(\zeta_{at,0}=-\infty\) and
\(\zeta_{at,K}=\infty\).
The design vector \(\bm x_q(X)\) excludes an intercept.
Let \(F_N\) and \(f_N\) denote the standard normal
distribution and density functions. Define
\[
c_{at,j}(X)=\zeta_{at,j}-\bm x_q(X)^\top\bm b_{at},
\qquad
q_{at}(k\mid X;\bm\theta_{at})
=F_N\{c_{at,k}(X)\}-F_N\{c_{at,k-1}(X)\},
\]
where
\(\bm\theta_{at}
=(\bm b_{at}^\top,\zeta_{at,1},\ldots,\zeta_{at,K-1})^\top\).
At the two boundary cutpoints, both \(f_N(c)\) and
\(cf_N(c)\) are interpreted as zero.

For \(j=1,\ldots,K-1\),
\[
\frac{\partial q_{at}(k\mid X;\bm\theta_{at})}
{\partial\bm b_{at}}
=\left[f_N\{c_{at,k-1}(X)\}
-f_N\{c_{at,k}(X)\}\right]\bm x_q(X),
\]
\[
\frac{\partial q_{at}(k\mid X;\bm\theta_{at})}
{\partial\zeta_{at,j}}
=I(j=k)f_N\{c_{at,k}(X)\}
-I(j=k-1)f_N\{c_{at,k-1}(X)\}.
\]
The Hessian blocks are
\[
\frac{\partial^2q_{at}(k\mid X;\bm\theta_{at})}
{\partial\bm b_{at}\partial\bm b_{at}^\top}
=\left[
c_{at,k-1}(X)f_N\{c_{at,k-1}(X)\}
-c_{at,k}(X)f_N\{c_{at,k}(X)\}
\right]\bm x_q(X)\bm x_q(X)^\top,
\]
\[
\begin{aligned}
&\frac{\partial^2q_{at}(k\mid X;\bm\theta_{at})}
{\partial\bm b_{at}\partial\zeta_{at,j}}\\
&={}\left[
I(j=k)c_{at,k}(X)f_N\{c_{at,k}(X)\}
-I(j=k-1)c_{at,k-1}(X)f_N\{c_{at,k-1}(X)\}
\right]\bm x_q(X),
\end{aligned}
\]
\[
\begin{aligned}
&\frac{\partial^2q_{at}(k\mid X;\bm\theta_{at})}
{\partial\zeta_{at,j}\partial\zeta_{at,l}}\\
&={} I(j=l)\left[
-I(j=k)c_{at,k}(X)f_N\{c_{at,k}(X)\}
+I(j=k-1)c_{at,k-1}(X)f_N\{c_{at,k-1}(X)\}
\right].
\end{aligned}
\]
For the observed category, the likelihood score is
\[
\bm s_{\theta_{at}}(O;\bm\theta_{at})
=\frac{I(A=a)}{q_{at}\{Y(t)\mid X;\bm\theta_{at}\}}
\frac{\partial q_{at}\{Y(t)\mid X;\bm\theta_{at}\}}
{\partial\bm\theta_{at}}.
\]
Its Jacobian is
\[
\begin{aligned}
\frac{\partial\bm s_{\theta_{at}}(O;\bm\theta_{at})}
{\partial\bm\theta_{at}^\top}
=I(A=a)\Bigg[
&\frac{1}{q_{at}\{Y(t)\mid X;\bm\theta_{at}\}}
\frac{\partial^2q_{at}\{Y(t)\mid X;\bm\theta_{at}\}}
{\partial\bm\theta_{at}\partial\bm\theta_{at}^\top}\\
&-\frac{1}{q_{at}\{Y(t)\mid X;\bm\theta_{at}\}^2}
\frac{\partial q_{at}\{Y(t)\mid X;\bm\theta_{at}\}}
{\partial\bm\theta_{at}}
\frac{\partial q_{at}\{Y(t)\mid X;\bm\theta_{at}\}}
{\partial\bm\theta_{at}^\top}
\Bigg].
\end{aligned}
\]
Averaging this Jacobian over the training sample gives
\(\widehat{\bm A}_{\theta_{at},v}\).

\paragraph{Multinomial logistic outcome model}
The initial multinomial model uses category \(1\) as
the reference. Set \(\bm b_{at,1}=\bm0\), and let
\(\bm x_q(X)\) include an intercept. Then
\[
q_{at}(k\mid X;\bm\theta_{at})
=\frac{\exp\{\bm x_q(X)^\top\bm b_{at,k}\}}
{\sum_{l=1}^K\exp\{\bm x_q(X)^\top\bm b_{at,l}\}},
\qquad
\bm\theta_{at}
=(\bm b_{at,2}^\top,\ldots,\bm b_{at,K}^\top)^\top.
\]
For \(j,l\in\{2,\ldots,K\}\),
\[
\frac{\partial q_{at}(k\mid X;\bm\theta_{at})}
{\partial\bm b_{at,j}}
=q_{at}(k\mid X;\bm\theta_{at})
\{I(k=j)-q_{at}(j\mid X;\bm\theta_{at})\}\bm x_q(X),
\]
\[
\bm s_{b_{at,j}}(O;\bm\theta_{at})
=I(A=a)\bm x_q(X)
\{I(Y(t)=j)-q_{at}(j\mid X;\bm\theta_{at})\},
\]
\[
\frac{\partial\bm s_{b_{at,j}}(O;\bm\theta_{at})}
{\partial\bm b_{at,l}^\top}
=-I(A=a)q_{at}(j\mid X;\bm\theta_{at})
\{I(j=l)-q_{at}(l\mid X;\bm\theta_{at})\}
\bm x_q(X)\bm x_q(X)^\top.
\]
Stacking these blocks in nonreference-category order
gives the probability derivative and nuisance score
Jacobian. The initial model uses reference category \(1\),
whereas the targeting fluctuation uses reference category \(K\).

\appendixsection{Complete results in real data analysis}

Table~\ref{tab:door-real-results} below presents the complete results for all five estimators in the real data analysis, whereas the main text reports only the results for TMLE-P.

\begin{table}[!htbp]
\centering
\small
\setstretch{1}
\renewcommand{\arraystretch}{1.05}
\caption{DOOR estimates using proportional-probit outcome models.}
\label{tab:door-real-results}

\begin{tabular*}{\textwidth}{@{\extracolsep{\fill}}cccccc@{}}
\toprule
Method & Visit (day) & Estimate & SE &
Pointwise 95\% CI & Simultaneous 95\% CI \\
\midrule
AIPW
 & 5       & 0.5951 & 0.0029 & (0.5894,\,0.6009) & (0.5848,\,0.6055) \\
 & 10      & 0.5118 & 0.0030 & (0.5060,\,0.5177) & (0.5013,\,0.5224) \\
 & 15      & 0.5016 & 0.0028 & (0.4961,\,0.5071) & (0.4917,\,0.5115) \\
 & 20      & 0.5002 & 0.0027 & (0.4949,\,0.5055) & (0.4906,\,0.5098) \\
 & 25      & 0.4993 & 0.0027 & (0.4941,\,0.5045) & (0.4899,\,0.5087) \\
 & 30      & 0.4985 & 0.0026 & (0.4933,\,0.5036) & (0.4892,\,0.5078) \\
 & Average & 0.5178 & 0.0024 & (0.5131,\,0.5224) & --- \\
 \hline
AIPW-CF
 & 5       & 0.5951 & 0.0029 & (0.5893,\,0.6008) & (0.5847,\,0.6055) \\
 & 10      & 0.5117 & 0.0030 & (0.5059,\,0.5176) & (0.5012,\,0.5223) \\
 & 15      & 0.5015 & 0.0028 & (0.4960,\,0.5070) & (0.4915,\,0.5114) \\
 & 20      & 0.5001 & 0.0027 & (0.4948,\,0.5055) & (0.4905,\,0.5098) \\
 & 25      & 0.4992 & 0.0027 & (0.4940,\,0.5044) & (0.4897,\,0.5087) \\
 & 30      & 0.4984 & 0.0026 & (0.4932,\,0.5035) & (0.4890,\,0.5077) \\
 & Average & 0.5177 & 0.0024 & (0.5130,\,0.5223) & --- \\
\addlinespace[4pt]
  \hline
TMLE
 & 5       & 0.5951 & 0.0029 & (0.5894,\,0.6008) & (0.5847,\,0.6055) \\
 & 10      & 0.5119 & 0.0030 & (0.5060,\,0.5177) & (0.5013,\,0.5224) \\
 & 15      & 0.5016 & 0.0028 & (0.4961,\,0.5071) & (0.4917,\,0.5115) \\
 & 20      & 0.5002 & 0.0027 & (0.4949,\,0.5055) & (0.4906,\,0.5098) \\
 & 25      & 0.4993 & 0.0027 & (0.4941,\,0.5045) & (0.4899,\,0.5088) \\
 & 30      & 0.4985 & 0.0026 & (0.4933,\,0.5036) & (0.4892,\,0.5078) \\
 & Average & 0.5178 & 0.0024 & (0.5131,\,0.5224) & --- \\
  \hline
TMLE-CF
 & 5       & 0.5950 & 0.0029 & (0.5893,\,0.6007) & (0.5846,\,0.6054) \\
 & 10      & 0.5116 & 0.0030 & (0.5058,\,0.5175) & (0.5011,\,0.5222) \\
 & 15      & 0.5014 & 0.0028 & (0.4959,\,0.5069) & (0.4915,\,0.5113) \\
 & 20      & 0.5001 & 0.0027 & (0.4947,\,0.5054) & (0.4904,\,0.5097) \\
 & 25      & 0.4991 & 0.0027 & (0.4939,\,0.5044) & (0.4897,\,0.5086) \\
 & 30      & 0.4983 & 0.0026 & (0.4932,\,0.5035) & (0.4890,\,0.5077) \\
 & Average & 0.5176 & 0.0024 & (0.5129,\,0.5223) & --- \\
  \hline
TMLE-P
 & 5       & 0.5950 & 0.0029 & (0.5893,\,0.6008) & (0.5846,\,0.6054) \\
 & 10      & 0.5117 & 0.0030 & (0.5059,\,0.5176) & (0.5012,\,0.5223) \\
 & 15      & 0.5015 & 0.0028 & (0.4960,\,0.5070) & (0.4915,\,0.5114) \\
 & 20      & 0.5001 & 0.0027 & (0.4948,\,0.5055) & (0.4905,\,0.5098) \\
 & 25      & 0.4992 & 0.0027 & (0.4940,\,0.5045) & (0.4898,\,0.5087) \\
 & 30      & 0.4984 & 0.0026 & (0.4932,\,0.5035) & (0.4890,\,0.5077) \\
 & Average & 0.5177 & 0.0024 & (0.5130,\,0.5223) & --- \\
\bottomrule
\end{tabular*}
\end{table}

\end{document}